\documentclass[journal]{IEEEtran}
\usepackage{amstext}
\usepackage{cite}
\usepackage{amsmath}
\usepackage{amsbsy}
\usepackage{amssymb}
\usepackage{mathtools}
\usepackage{graphicx}
\usepackage[font={small},labelfont={bf}]{caption}
\usepackage{subcaption}
\usepackage{multirow}
\usepackage{booktabs}
\usepackage{soul}
\usepackage{algorithm,caption}
\usepackage{pifont}
\usepackage{xcolor}
\DeclareUnicodeCharacter{202A}{~}

\begin{document}
%
\title{Driver Profiling and Bayesian Workload Estimation Using Naturalistic Peripheral Detection Study Data}

%

\author{Nermin Caber$^\dagger$,
        ~Bashar I. Ahmad$^\dagger$,
        ~Jiaming Liang$^\dagger$,
        ~Simon Godsill,
        ~Alexandra Bremers,
        ~Philip Thomas,
        ~David Oxtoby
        ~and ‪Lee Skrypchuk
\thanks{N. Caber, B. I. Ahmad, J. Liang and S. Godsill are with the Department
of Engineering, University of Cambridge, Cambridge, CB2 1PZ, UK, Emails: \{nc501, jl809, bia23, sjg30\}@cam.ac.uk.}
\thanks{A. Bremers is with Cornell Tech, NY, USA. Email:awb227@cornell.edu.}
\thanks{P. Thomas, D. Oxtoby and L.  Skrypchuk are with Jaguar Land Rover, Whitley, UK. Emails:\{pthoma24, doxtoby, lskrypch\}@jaguarlandrover.com.}
\thanks{$^\dagger$Authors made an equal contribution to this work.}
}


\maketitle

\begin{abstract}
Monitoring drivers' mental workload facilitates initiating and maintaining safe interactions with in-vehicle information systems, and thus crucial for delivering adaptive human machine interaction solutions with reduced impact on the primary task of driving. In this paper, we tackle the problem of workload estimation from driving performance data. First, we present a novel on-road study for collecting subjective workload data via a modified peripheral detection task in naturalistic settings. Key environmental factors that induce a high mental workload are identified via video analysis, e.g. junctions and behaviour of vehicle in front. Second, a supervised learning framework using state-of-the-art time series classifiers (e.g. convolutional neural network and transform techniques) is introduced to profile drivers based on the average workload they experience during a journey. A Bayesian filtering approach is then proposed for sequentially estimating, in (near) real-time, the driver's instantaneous workload. This computationally efficient and flexible  method can be easily personalised to a driver (e.g. incorporate their inferred average workload profile), adapted to driving/environmental contexts (e.g. road type) and extended with data streams from new sources. The efficacy of the presented profiling and instantaneous workload estimation approaches are demonstrated using the on-road study data, showing $F_{1}$ scores of up to 92\% and 81\%, respectively.
\end{abstract}


\IEEEpeerreviewmaketitle

\section{Introduction}
\IEEEPARstart{E}{very} year 1.35 million people die in road traffic accidents according to the World Health Organization \cite{world2018global}; this makes it the main cause of death for people aged 15-29 years. The majority of these accidents can be attributed to human factors, such as distraction~\cite{department2021reported}. This is exacerbated by modern In-Vehicle Information Systems (IVIS) of growing complexities, such as navigation and driver assistance systems \cite{harvey2013usability}. Their use can increase drivers' workload and thus affect their ability to carry out the primary task of driving. Hence, adaptive Human-Machine Interaction (HMI) has emerged as a solution to improve the usability of IVIS by delivering information and/or initiating-maintaining an interaction with drivers only at ``appropriate'' times based on their current mental workload \cite{bellotti2005comunicar,amditis2010towards,yi2020implicit, gomaa2022s}. The  measurement and estimation of drivers' workload also notably benefited from recent advances in sensing technology and Machine Learning (ML) algorithms, e.g.~\cite{fridman2018cognitive, semmens2019now, yi2019new, cha2020hello, gomaa2022s}.

For user safety, acceptance and satisfaction reasons, drivers subjectively reporting the appropriate times an adaptive HMI system could interact with them is a practical means to capture representative data of the cognitive workload in naturalistic driving scenarios (see related work in Section \ref{sec:relatedwork} below). This approach is adopted in this paper since it circumvents the need for devising objective criteria for such suitable time intervals, which can be prohibitively complex given their dependence on various nuances such as contextual factors not always easily observable in on-road studies \cite{fridman2018cognitive}. In particular, the following two forms of subjective mental workload are defined here: 
\begin{enumerate}
    \item \textit{Instantaneous Workload Level (IWL)}: is the current workload experienced by the driver at any time instant as in other related studies, e.g. \cite{de1996measurement, wright2017intelligent,fridman2018cognitive,de2019can, semmens2019now}. 
    \item \textit{Average Workload Profile (AWP)}: represents the driver's long-term workload during a journey. This can reflect the user's individual workload management capabilities where some drivers on average and over prolonged periods sustain lower workload than others under identical conditions. This can be due to their prior driving experience, which 
    can reduce the effort of safely performing the vehicle control task \cite{de1996measurement,rasmussen1986information}. This is supported by \cite{yi2019personalized,yi2019new} who highlight the various workload categories explored by different drivers. AWP is assumed to be stable and distinct from drivers' responding to environmental-contextual factors that trigger high IWLs.
\end{enumerate}
\vspace{-4mm}
\subsection{Problem Statement}\label{sec:problemstatement}
The overall objective in this paper is to enable adaptive HMI via the robust estimation of the driver's \textit{subjective} workload from easily accessible and widely available data sources in vehicles (e.g. driving performance signals).
For this purpose, we address here the following research questions: 
\begin{itemize}
   \item RQ1: What environmental and driving contexts induce a high IWL in naturalistic settings? 
   \item RQ2: Can we cluster drivers into different groups as per their observed AWP during a journey? Specifically, 
    \begin{itemize}
        \item RQ2.1: How many AWP groups can be identified and how do they relate to subjective driving styles? 
        \item RQ2.2: How can a driver be matched to any of the AWP categories identified from driving performance data using supervised learning? 
    \end{itemize}
    \item RQ3: How can the IWL be estimated in (near) real-time from asynchronous data sources, incorporating specific driver attributes (e.g. AWP) and/or contextual information (e.g. road type, traffic, etc.)?
\end{itemize}

\subsection{Proposed Approach and Contributions} \label{sec:contributions}
We present in this paper an on-road study with $N=24$ participants that was specifically designed to answer the above research questions. The main contributions are:
\begin{itemize}
\item A novel study design with a modified Peripheral Detection Task (PDT).
\item Identifying via video analysis external factors that contribute to high instantaneous workload levels in naturalistic driving scenarios (e.g. junctions and vehicles in front).
\item Ascertaining from the collect on-road data that drivers can have notably different average workload profiles and thereby confirming the need for personalised instantaneous workload estimators; no clear correlation between AWP and self-reported driving styles is also established.
\item Proposing a supervised learning framework to determine a driver's AWP from limited training data
.
\item
Introducing a simple, yet effective, Bayesian filtering approach to \textit{sequentially} track the driver's IWL from asynchronous data streams (e.g. CAN-bus signals); it can be easily personalised (e.g. leveraging the driver's AWP) and adapted to new data sources (e.g.  additional data sources or contextual information such as the road type).
\end{itemize}
The study data is used to demonstrate the efficacy of the proposed AWP and IWL estimation techniques.

\subsection{Paper Layout}
In Section \ref{sec:relatedwork} key related work and its distinction from the approach presented in this paper is outlined. Whilst the experimental study and collected measures are described in Section \ref{sec:experimentalstudy}, results of the video analysis and AWP along with its correlation with driving style data are presented in Section \ref{sec:videoandquestionnair}. A classification framework for inferring the driver's AWP is discussed in Section \ref{sec:AWPestimation}. The Bayesian approach for instantaneous workload estimation is detailed and evaluated in Section \ref{sec:workloadestimation}. Conclusions are drawn in Section \ref{sec:conclusions}.

\section{Related Work}\label{sec:relatedwork}



%
Adaptive HMIs which adjust the user interface (e.g. suppress a phone call) as a response to some stimuli such as a challenging driving environment have a long history in the automotive sector~\cite{yi2020implicit}. For instance, the GIDS project explored tailoring information to avoid mental overload and deliver a highly personalised driver experience in the 1990s~\cite{michon1993human}. It applied a rule-based method to estimate the driver's mental workload. 
Many subsequent projects, e.g. CEMVOCAS~\cite{bellet2007available}, COMUNICAR~\cite{bellotti2005comunicar}, and AIDE~\cite{amditis2010towards}, as well as independent studies, e.g. ~\cite{fridman2018cognitive, yi2020implicit,de2019can}, pursued similar goals with data-driven inference approaches. Notably, the AIDE project demonstrated through multiple on-road studies that drivers prefer adaptive HMIs over static ones~\cite{amditis2010towards}. In this paper and similar to recent work on driver workload estimation, such as ~\cite{fridman2018cognitive, semmens2019now, de2019can, yi2020implicit, gomaa2022s}, data-driven inference is adopted, capitalising on advances in ML and Bayesian filtering, contrary to applying rule-based methods, such as \cite{michon1993human,wright2017intelligent}.

Estimating the driver workload from physiological information (e.g. ocular measures and electrocardiac activity) is particularly popular, e.g. see \cite{mehler2012sensitivity, solovey2014classifying, fridman2018cognitive,yi2019new, de2019can, xie2019personalized, gomaa2022s}, due to known linkages of such data to mental workload and the ability of associated sensors to provide continuous, online observations~\cite{de1996measurement}. On the other hand, there is a growing interest, e.g. see \cite{yi2019personalized, wright2017intelligent , semmens2019now, de2019can, yi2019new}, in relying solely on non-intrusive driving performance signals (e.g. steering, braking, and speed) since they are often widely available and easily accessible in modern vehicles, via the CAN bus. They can also deliver accurate IWL estimates \cite{de2019can,yi2019new} and for these reasons this approach is adopted here. It circumvents the common limitations of using physiological data in naturalistic driving scenarios, such as the necessity to establish a driver's baseline~\cite{de1996measurement} and employing specialised sensor instrumentation as well as signal processing~\cite{mehler2012sensitivity, young2015state, de2019can, gomaa2022s}, with potentially high adoption costs in mainstream vehicles. However, physiological observations, if/when available, can be utilised, e.g. within the presented IWL inference technique.   

The sought driver IWL in many studies is specifically that induced by a secondary task~\cite{fridman2018cognitive,mehler2012sensitivity,reimer2012field} to assess its impact on safety, e.g. \cite{mehler2012sensitivity} introduces a delayed digit recall task (i.e. n-back) whose difficulty level is considered as the workload ground truth. Conversely, in \cite{schneegass2013data, cnossen2000strategic} data on the primary task is collected and analysed to establish how the driving environment influences drivers' IWLs. For instance, it is reported that motorways induce lower workload than urban roads \cite{schneegass2013data} and rushed driving increases experienced workload \cite{cnossen2000strategic}. 
In such settings, post-hoc video rating of the route where drivers subjectively evaluate their workload on a pre-defined scale is a reasonable means to establish the ground-truth IWL~\cite{schneegass2013data}, with the main disadvantage of not being dynamic and requiring drivers to recall the specific driving situations. Nevertheless, many researchers consider subjective evaluation to be the best measurement technique as the affected individual should be able to evaluate their mental workload most accurately~\cite{de1996measurement,wickens2020processing}. In this paper and similar to~\cite{cnossen2000strategic,schneegass2013data}, we consider workload induced (predominantly) by the primary vehicle control task and design an experiment to capture data for developing IWL estimation algorithms. However, in this work we adopt a novel means, i.e. a modified version of the Peripheral Detection Task (PDT), to collect subjective workload information during driving to overcome limitations of post-hoc video rating.

In \cite{semmens2019now} and \cite{cha2020hello} subjective data is collected in naturalistic automotive and domestic contexts respectively, namely by an experimenter repeatedly asking the subject, e.g. a driver \cite{semmens2019now}, ''Is now a good time?''. Here, we propose applying a modified version of the PDT, which automates the data acquisition and supports higher frequency of self-reporting compared to that in \cite{semmens2019now, cha2020hello}. PDT is  a continuous workload measurement, which has proved sensitive to workload variations~\cite{martens2001effects,rupp2010peripheral,schaap2008drivers}. The traditional PDT dictates that drivers respond to a peripheral, visual stimuli by pressing a finger-worn button. In this paper, the peripheral, visual stimuli replaces the auditory verbal message (i.e., ``Is now a good time?'') in \cite{semmens2019now} and conveys the question ``Is your instantaneous workload low?''. This leads to more (training) data being available to devise workload estimation algorithms, thereby potentially improving accuracy and robustness of any applied data-centric inference method.

The majority of estimators for instantaneous mental workload, e.g. in\cite{amditis2010towards, solovey2014classifying, fridman2018cognitive, semmens2019now, yi2019new, yi2020implicit, cha2020hello, gomaa2022s}, are based on established ML algorithms such as Support Vector Machines (SVM), and reliable neural network models.  Whereas, in this paper we propose a filtering approach that applies the Bayesian recursion, as is common in the object tracking field \cite{sarkka2013bayesian}, based on a predefined Markov chain (MC) model and an empirically learnt observation likelihood. This is also distinct from common uses of Hidden Markov Models (HMMs) learnt directly from data, e.g. as in \cite{fridman2017can}. Here, the MC transition matrix values are physically meaningful and can be intuitively set, thus reducing the training data requirements. The introduced probabilistic approach sequentially estimates the posterior distribution of the IWL, thereby capturing the inference uncertainty over time unlike rule-/SVM-based or clustering methods, e.g. \cite{wright2017intelligent, semmens2019now, gomaa2022s}. It can also naturally treat asynchronous data from multiple sources and easily incorporate additional information such as the user profile (e.g. via adapting the MC matrix) and even new data sources (e.g. via updating the likelihood distribution). Such modifications do not require complete retraining of the estimator unlike with standard purely data-driven techniques that demand synchronised data at a fixed rate from all sources. Their application with asynchronous CAN-bus signals typically necessitates ad-hoc data formatting, e.g. averaging within a finite time window as in \cite{wright2017intelligent, semmens2019now} which can inadvertently conceal salient features in the raw signals and introduce processing latencies.

Finally, in this work we address the problem of inferring the driver's AWP, rather than only IWL as in prior work such as \cite{amditis2010towards, solovey2014classifying, fridman2018cognitive, semmens2019now, cha2020hello}. This is motivated by~\cite{yi2019new} who show the existence of clusters/groups of drivers within (instantaneous) workload data and the importance of such information for personalising workload estimation models, e.g. to circumvent the need to have sufficient data for a specific (new) user. However, unlike~\cite{yi2019new, xie2019personalized} where biometric signals and conventional ML clustering/recognition methods are used, a personalisable Bayesian filtering approach is introduced here for IWL estimation from driving performance data streams. Whilst we also apply state-of-the-art time series classification algorithms \cite{ismail2020inceptiontime,dempster2021minirocket} for profiling,  the limited available training data necessitates utilising short input sequences and combining AWP inference results from separate data segments in a journey. This is presented here within a framework, including a discussion on the performance metrics. We also demonstrate how AWP can be easily leveraged to deliver a personalised IWL estimator with improved accuracy.

\section{Experimental Study}\label{sec:experimentalstudy}
\begin{figure}[t]
	\centering    
	\includegraphics[width=8.75cm,height=6.5cm]{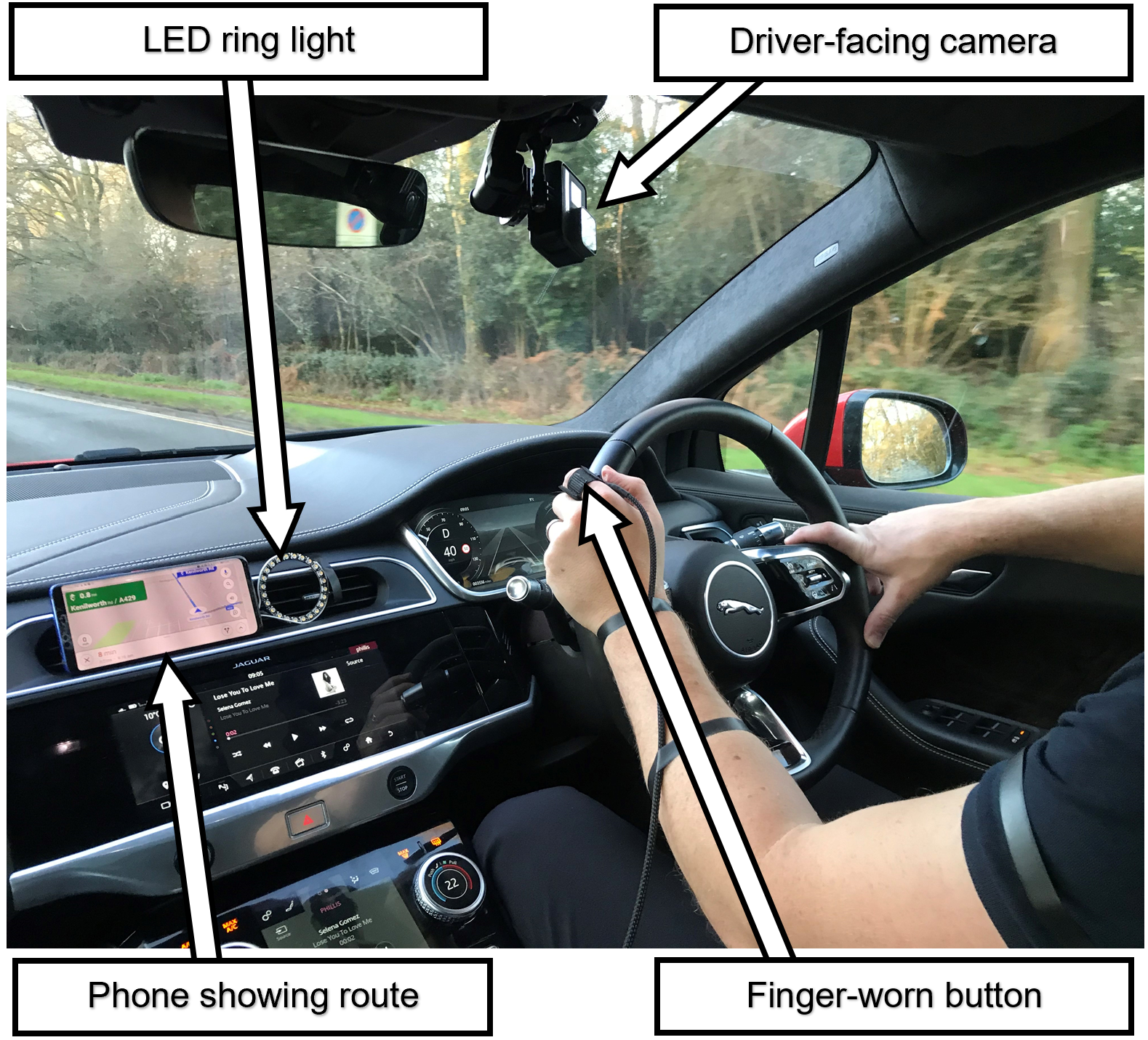}
	\caption[Experimental setup of the study vehicle interior.]{Experimental setup of the study vehicle interior.}
	\label{fig:Instrumented_Study_Vehicle}
\end{figure}

The experimental car, provided by Jaguar Land Rover (JLR), was a battery electric vehicle, namely a Jaguar I-PACE EV400 AWD. A phone was mounted onto the central air vent, a typical space for navigation systems, and showed the route the participant had to follow (see Figure~\ref{fig:Instrumented_Study_Vehicle}). Next to the phone, an LED ring light was installed which lit up in red every 5 to 10 seconds, resembling the PDT~\cite{martens2001effects} which uses an interval of 3 to 5 seconds. This red light was intended as a subtle and quick way to ask participants about their workload. 

Participants were equipped with a small finger-worn button (see Figure~\ref{fig:Instrumented_Study_Vehicle}) which they were asked to press in perceived low-workload situations when the LED ring light is lit in red. This setup deviates slightly from the traditional PDT since here the LED was a reminder (i.e. prompt) to the drivers to report their subjective instantaneous workload. Thus, it represented more a hint than a task to be followed as is the case in the original PDT~\cite{martens2001effects}. This was also the reason for decreasing the frequency in comparison to~\cite{martens2001effects} since here drivers had to cognitively process the request (i.e. evaluate their workload) instead of just reacting to the lit up LED. 

\vspace{-2mm}

\subsection{Participants}
All participants were recruited internally at JLR, had valid driver licences and had undergone an internal safety driving course beforehand. Although all participants were familiar with JLR cars, not all of them had experienced an electric Jaguar I-PACE before. In total, $24$ drivers ($4$ females) ranging from $27$ to $58$ years old ($M=39.8$, $SD=8.2$) participated in the study. Their driving experience ranged between $4$ and $40$ years ($M=21.2$, $SD=9.4$). None of the participants had impeding motor-visual impairments, including colour blindness. No compensation was offered for their participation. Participants' informed consent was obtained prior to the study. All data were collected between January and March 2020.

\subsection{Study Design and Collected Measures} \label{sec:collectedmeasures}
\begin{figure}[b] 
	\centering    
	\includegraphics[width=8.5cm,height=7.5cm]{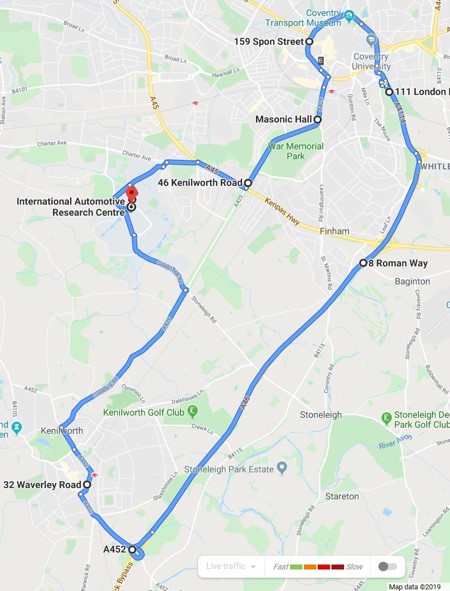}
	\caption[Map depicting the experimental route.]{Map of the experimental route followed by participants.}
	\label{fig:Study_Route}
\end{figure}

The study was designed as an on-road driving experiment to increase the ecological validity of the collected driving data. All participants started the experiment at 10am to ensure similar traffic and undertook two successive drives:
\begin{enumerate}
    \item \textbf{Familiarisation route}: The route was approximately 22 miles long, took 40 minutes, and consisted of rural, urban and main roads. 
    This drive acclimatised drivers with the car, the HMI and the task. First, drivers had to become comfortable with the unfamiliar driving characteristics of an electric car, i.e. the very responsive accelerator and the strong regenerative braking when coasting. To facilitate this familiarisation, the car's regenerative braking mode was set to closely imitate a petrol car's coasting behaviour. Second, drivers familiarised themselves with the car's and the experiment's HMI, e.g. the operation of the finger-worn button. Finally, drivers had to become comfortable with the task of affirming with a button press the question ``Is your instantaneous workload low?'' which was represented by the lit-up, red LED ring. Due to the relatively high frequency of requests, the experimenter ensured that the participants learned to press the button only when they perceived the driving situation as low workload. This was achieved during the familiarisation run by repeatedly pointing out that the red LED ring light should not be considered a task and interpreted as a trigger to press the button. No data were collected during the familiarisation drive.
    \item \textbf{Experimental route}: The experimental route (see Figure~\ref{fig:Study_Route}) incorporated the UK road types: i) major (e.g. A roads for large-scale transport linking regional towns and cities), ii) B roads (connect different areas and feed traffic between A roads and smaller roads) and iii) classified-unnumbered/unclassified (smaller roads for connecting A and B roads as well as for local traffic). It covered a distance of approximately 18 miles, and lasted around 40 minutes. The various road types ensured that the collected driving data were as naturalistic as possible and covered a wide range of different environments.
\end{enumerate}


The drivers' performance measures, i.e. dependent variables, were collected from the following range of sources:
\begin{itemize}
    \item \textbf{Driving Style Questionnaire (DSQ)}: data from a standardized questionnaire~\cite{french1993decision} that captures a driver's self-evaluated driving style using six dimensions, i.e. speed, calmness, social resistance, focus, planning, and deviance; it was completed prior to the familiarisation drive, see the procedure below (Section~\ref{subsec_procedure}).
    \item \textbf{Controller Area Network (CAN) bus}: many signals related to the driving performance. Their data arrives asynchronously at an overall rate of $\approx200$Hz with example streams listed in Table~\ref{table:CAN_Bus_OnRoad}, all of which, except for the GPS ones, are used for estimating the driver's AWP and IWL below. These signals were selected because they can be easily extract from the raw CAN-bus data and are utilised in related studies, e.g.  \cite{yi2019personalized, wright2017intelligent , semmens2019now, de2019can, yi2019new}. 
    \item \textbf{Visual prompts and button presses}: 
    time instants and duration of visual prompts (i.e. LED ring lit in red) and (finger-worn) button presses to record the drivers' IWL.
    \item \textbf{Video streams}: from two GoPro cameras, forward and driver facing, to analyse environmental factors influencing the reported IWL, and monitor the cabin status (e.g. to verify the experimental procedure).
\end{itemize}

\begin{table}
	\caption{Description of selected CAN bus data.}
	\centering
	\label{table:CAN_Bus_OnRoad}
	\begin{tabular}{p{0.3\columnwidth} p{0.6\columnwidth}}
		\toprule
		CAN bus data  & Description \\ 
		\midrule
		
		
		GPS\textunderscore Latitude & Latitude GPS coordinate of the car \\
		
		GPS\textunderscore Longitude & Longitude GPS coordinate of the car \\
		
		VehicleSpeed & Speed of the car in mph \\
		
		SteeringWheelAngle & Angle of the steering wheel in the range $-780$ to $780$ deg \\ 
		
		SteeringWheelAngleSpeed & Angular velocity of the steering wheel in the range 0 to 1016 $\frac{deg}{s}$ \\
		
		PedalPos & Position of the accelerator in the range $0$ (starting position) to $100$ (end position) \\
		
		BrakePressure & Brake pressure in the range $0$ to $204.6$ $Bar$ (indicating brake pedal position) \\
		
		LateralAcceleration & Acceleration perpendicular to the direction of travel in the range $-11$ to $11$ $\frac{m}{s^{2}}$ \\
		
		YawRate & Angular velocity of the car's rotation around the vertical axis in the range $-100$ to $100$ $\frac{deg}{s}$ \\
		
		
		\bottomrule
	\end{tabular}
\end{table}
\vspace{-3mm}
\subsection{Procedure} \label{subsec_procedure}
The experiment started at a JLR parking garage where participants were briefed on the study's goal and procedure, including the routes. Subsequently, they completed the required documentation, consisting of a consent form, demographic sheet and driving style questionnaire~\cite{french1993decision,west1993direct}. In the car, participants were introduced to the vehicle's specifications, its HMI and the experimental equipment. It was emphasised that the purpose of the lit LED ring light was to ask participants ``Is your instantaneous workload low?'' They were instructed to press the finger-worn button to answer ``Yes'' and to refrain from pressing to answer ``No''. Participants were also told to comply with the highway code at all times.

After the experimenter activated the LED ring light, participants drove off and followed the familiarization route. 
Participants then had a break of approximately $60$ minutes, which gave them a chance to ask questions and rest.

Upon the participant's return from the break, the experimenter activated the LED ring, the CAN bus logging and the two cameras, and set up the experimental route on the phone. Subsequently, drivers drove off and completed the $40$-minute-long experimental route. Despite the navigation system on the phone, a few drivers went off route and had to be rerouted.
\vspace{-3mm}
\subsection{Instantaneous Workload Labelling and Average Profiles}
\begin{figure}[t] 
	\centering    
	\includegraphics[width=8.75cm,height=4.5cm]{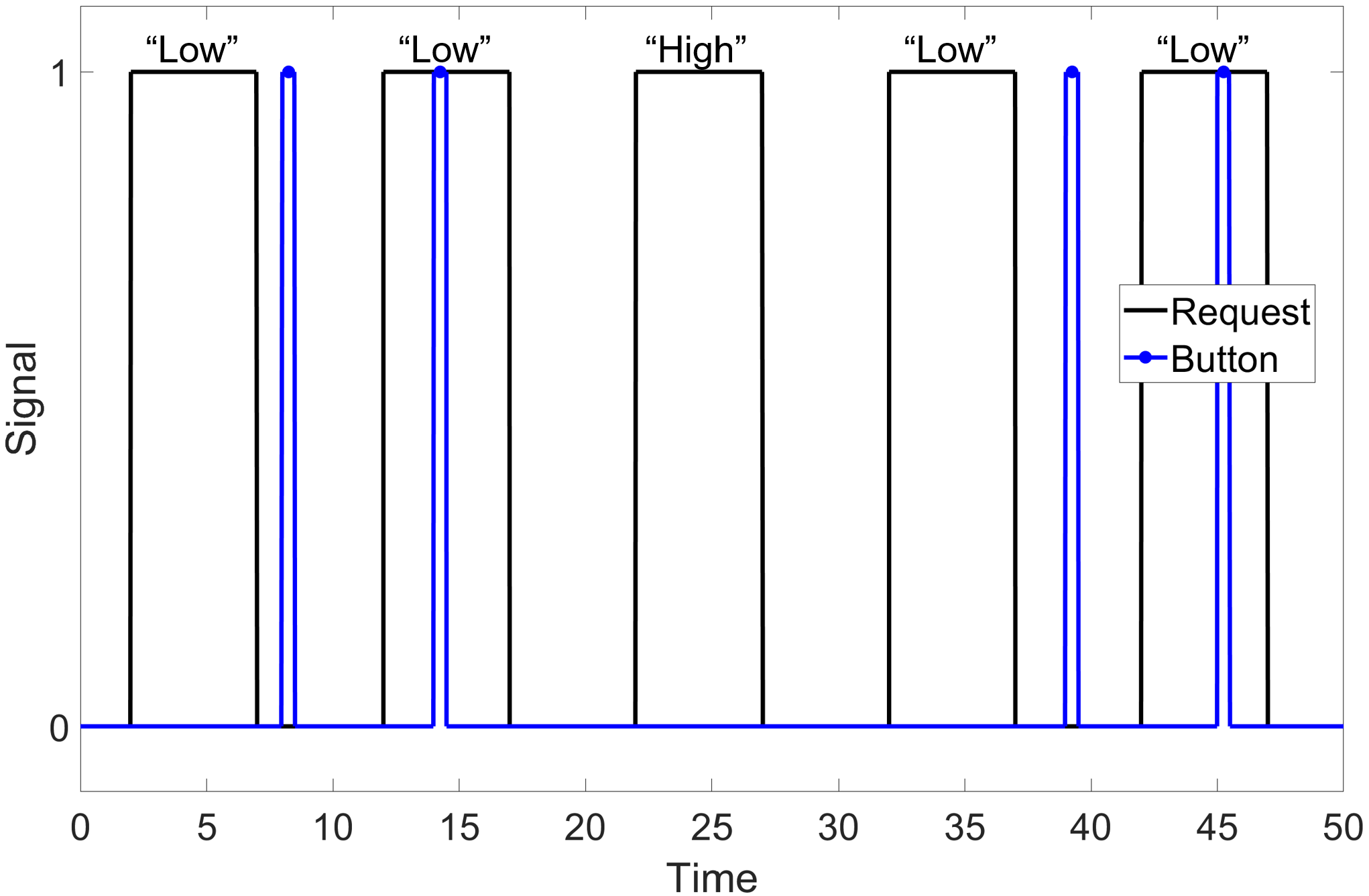}
	\caption[Plot depicting request and button signals to illustrate the labelling of instantaneous workload.]{
	Example requests (visual prompts) and button-press signals to illustrate the IWL labels, low and high.}
	\label{fig:WorkloadLabelling}
\end{figure}

All 24 participants completed the experimental route, with a total of 3730 button presses, 5264 requests, and 828 minutes of driving performance and video data recorded. The workload data are labelled as either ``High'' or ``Low'' as in the standard PDT~\cite{martens2001effects}. In particular, in this paper it is considered ``Low'' when a button press occurred during or after the request (i.e. visual prompt), and conversely ``High'' when no button press occurred for a specific request. An illustrative example is depicted in Figure~\ref{fig:WorkloadLabelling}. This simple labelling strategy is motivated by adaptive HMI requirements, e.g. revealing appropriate interaction times, and the binary nature of the drivers' reports. 

Additionally, the average workload experienced by a participant throughout a drive is captured here by the Low Workload Ratio (LWR), $\text{LWR} \in [0,1]$, as per
\begin{align} \label{eq:LWR}
	\text{LWR} = \frac{\#Low}{\#Low+\#High},
\end{align} 
where $\#X$ is the number of occurrences of event $X$ in a journey completed by a participant during the experimental drive. It is the fraction of situations that a participant reported a low IWL during the experimental drive; a low LWR score signifies a high AWP and vice versa (see  Section \ref{sec:AWPanalysis}).

\vspace{-3mm}
\subsection{Limitations of the Study}
Various limitations need to be considered when interpreting the study results. First, external factors, such as weather and traffic, can influence a driver's self-evaluated workload and consequently their grouping. As the video analysis did not reveal a strong influence of some external factors and all experiments were carried out at approximately the same time, ensuring similar traffic levels, the overall effect should be small, albeit present. Second, there can be a dissociation between drivers' subjective workload assessment and their actual performance on a specific task~\cite{horrey2009dissociation}. It is challenging to confirm that the self-reported status (including with PDT) corresponds to the driver's true mental/cognitive workload (e.g. due to lack of familiarity with the PDT procedure) without collecting reliable physiological data with baselines. Despite this limitation, self-evaluated workload is an accurate source of data~\cite{de1996measurement,wickens2020processing} and has been used widely in human factors research~\cite{schaap2008drivers, schneegass2013data, semmens2019now}. Third, all participants were JLR employees (only four were females) and OEM employees can deviate from the average driver in several aspects, such as automotive knowledge and experience with car interfaces. Fourth, whilst in this work the driver's IWL is either high or low, in practice it is not binary and can even be treated as a continuous variable, e.g. \cite{wright2017intelligent}. The adopted simplified labelling here can nonetheless be viewed as reasonable in the context of adaptive HMI, especially given the binary nature of the available (ground-truth) subjective IWL reporting. More elaborate labelling schemes can be considered in future work, for instance based on the duration between the prompt and pressing the button as well as the frequency of successive missed presses (e.g. four labels with medium low and medium high IWLs). Finally, the choice of red color for the visual prompts followed from the PDT study \cite{martens2001effects}. Its potential impact on the self-reported IWL is unknown as colours can affect humans' behaviour~\cite{elliot2014color}.

\section {Factors and Average Workload Profiles} \label{sec:videoandquestionnair}
\subsection{Video Analysis of High Instantaneous Workload (RQ1)}
The forward-facing videos were studied to understand what environmental factors triggered high instantaneous workload. This is called critical incident analysis~\cite{flanagan1954critical,day2018drivers} and is typically guided by a study question~\cite{mackenzie1999video}, i.e. RQ1 in Section~\ref{sec:problemstatement}. Here, analysis started $10$ seconds before each reported high IWL to comprehend the driving environment.

\begin{figure}[t] 
	\centering    
	\includegraphics[width=0.7\columnwidth]{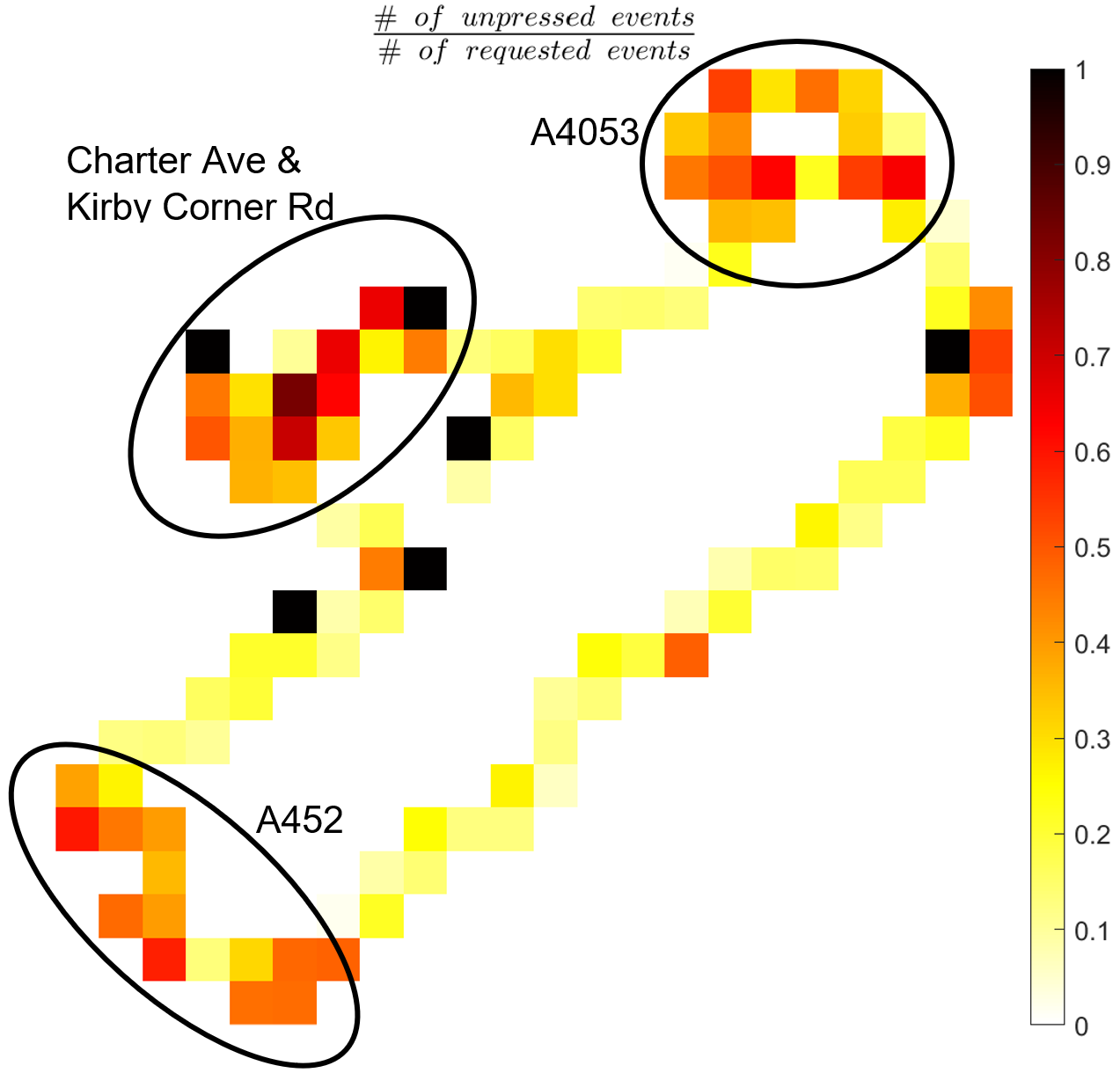}
		\caption[Heatmap illustrating the road sections' varying impact on drivers' perceived workload.]{Heat map of recorded $1-\text{LWR}$ (i.e. high IWL incidents) within discretised parts of the route for all $24$ participants.}
	\label{fig:HeatMap}
\end{figure}

In total, 12 factors (see Table~\ref{table:Video_Workload}) were identified from 1534 incidents of high workload. Since some high IWL events could not be attributed to a cause, the percentages in Table~\ref{table:Video_Workload} do not add up to 100\%. Notably, ``junctions'' and the ``car in front'' account for approximately 53\% of all self-evaluated high workload scenarios. This is unlike generic high traffic density. Noticeable differences in perceived workload were also observed depending on the road type. This is based on plotting $1-\text{LWR}$, i.e. fraction of number of unpressed events signifying high IWLs,  within a predefined grid of the experimental drive route; see Figure~\ref{fig:HeatMap} for all $24$ participants. As expected, the urban parts (including junctions) induced higher workload than other road sections for most drivers. 

\begin{table}
	\caption[Identified environmental factors causing high workload based on video analysis.]{Identified environmental factors causing high workload based on the conducted video analysis.} 
	\centering
	\label{table:Video_Workload}
	\begin{tabular}{l c}
		\toprule
		Environmental Factor & Percentage of incidents \\ 
		\midrule
		Approaching, observing and turning at junction & 39\% \\
		
		Behaviour of vehicle in front (e.g. braking) & 14\% \\
		
		Changing lanes (particularly motorway) & 7\% \\
		
		Vulnerable road users (e.g. pedestrians) & 7\% \\
		
		Narrow road (sometimes with oncoming traffic) & 6\% \\
		
		High traffic density & 5\% \\
		
		Accelerating to desired speed & 4\% \\
		
		Obstacles on the road (e.g. parked car) & 4\% \\
		
		Reading traffic sign & 3\% \\
		
		Roadworks & 2\% \\
		
		Road bend & 2\% \\
		
		Joining motorway & 2\% \\
		\bottomrule
	\end{tabular}
\end{table}
\vspace{-3mm}
\subsection{Average Workload Profile (RQ2)}\label{sec:AWPanalysis}
\begin{figure}[t] 
	\centering    
	\includegraphics[width=1\columnwidth]{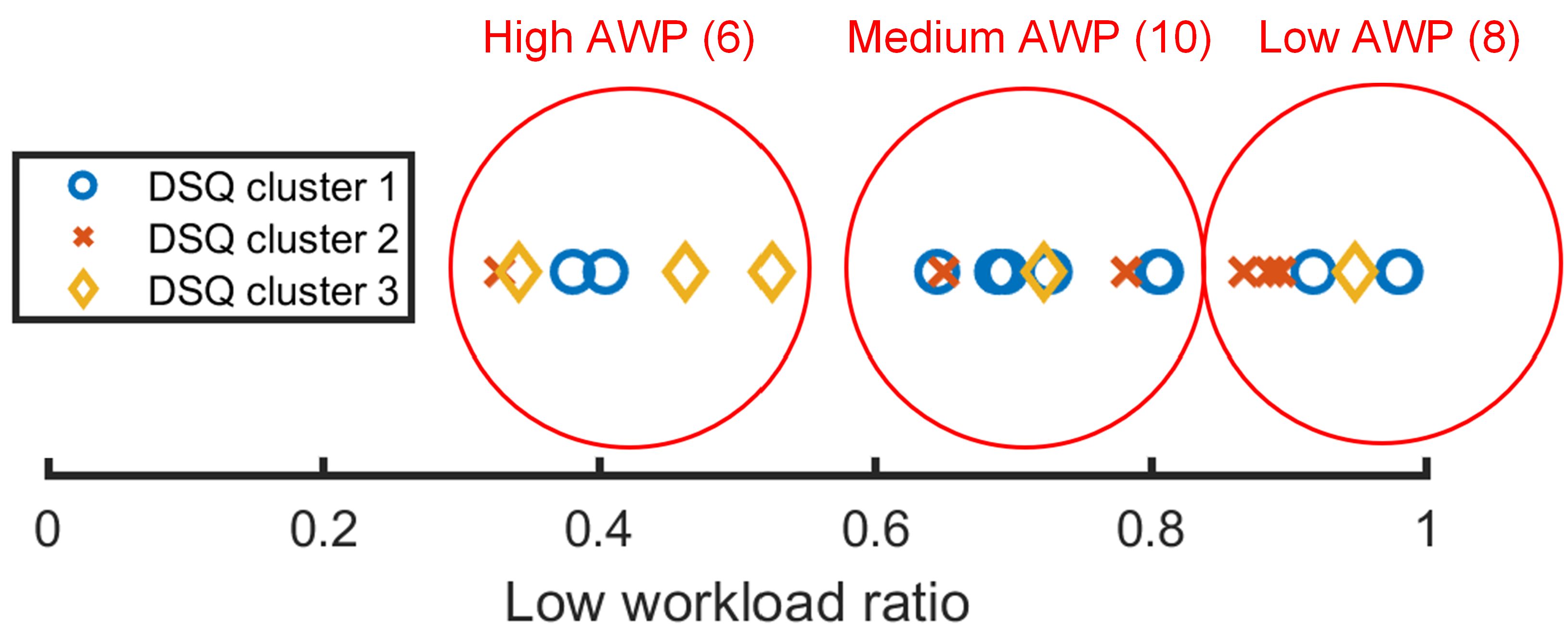}
	\caption[Scatter plot of participants' DSQ clusters and their AWP including average workload profiles.]
	{Scatter plot of the low workload ratio in (\ref{eq:LWR}) for all participants. Red circle are for three identified groups/clusters (number of drivers per group is in the brackets). Different shapes (i.e. circle, cross, and diamond) depict the DSQ cluster associated with each driver.}
	\label{fig:WorkloadDSQ}
\end{figure}

The frequency of reporting high IWL drastically varied between the participants during the experimental drive. This is visible in the scatter plot in Figure~\ref{fig:WorkloadDSQ}. It depicts each participant's LWR in (\ref{eq:LWR}), with values ranging from $32.81\%$ to $98.05\%$. Visual inspection of this plot clearly reveals three potential groups of drivers, each with a distinctive profile in relation to average experienced workload during a journey (k-means algorithm also confirmed this clustering). These groups are (i.e. as per their LWR score): low (34\%-52\%), medium (64\%-81\%) and high (87\%-98\%). Accordingly,  three average workload profiles of drivers, i.e. $\text{AWP}\in\{L,M,H\}$, are defined in this work as per 
\begin{equation} \label{eq:AWPDefinition}
   \text{AWP}= \left\{ 
  \begin{array}{ c l }
    H & \quad \text{if~~~~} \text{LWR} \leq 55\% \\
    M & \quad \text{if~~~~} 55\% < \text{LWR} \leq 85\% \\
    L & \quad \text{if~~~~}\text{LWR} > 85\% \\
  \end{array}
\right.
\end{equation}

\begin{figure}[t]
\centering
\begin{subfigure}[t]{0.492\linewidth}
\includegraphics[width=4.45cm,height=4.45cm]{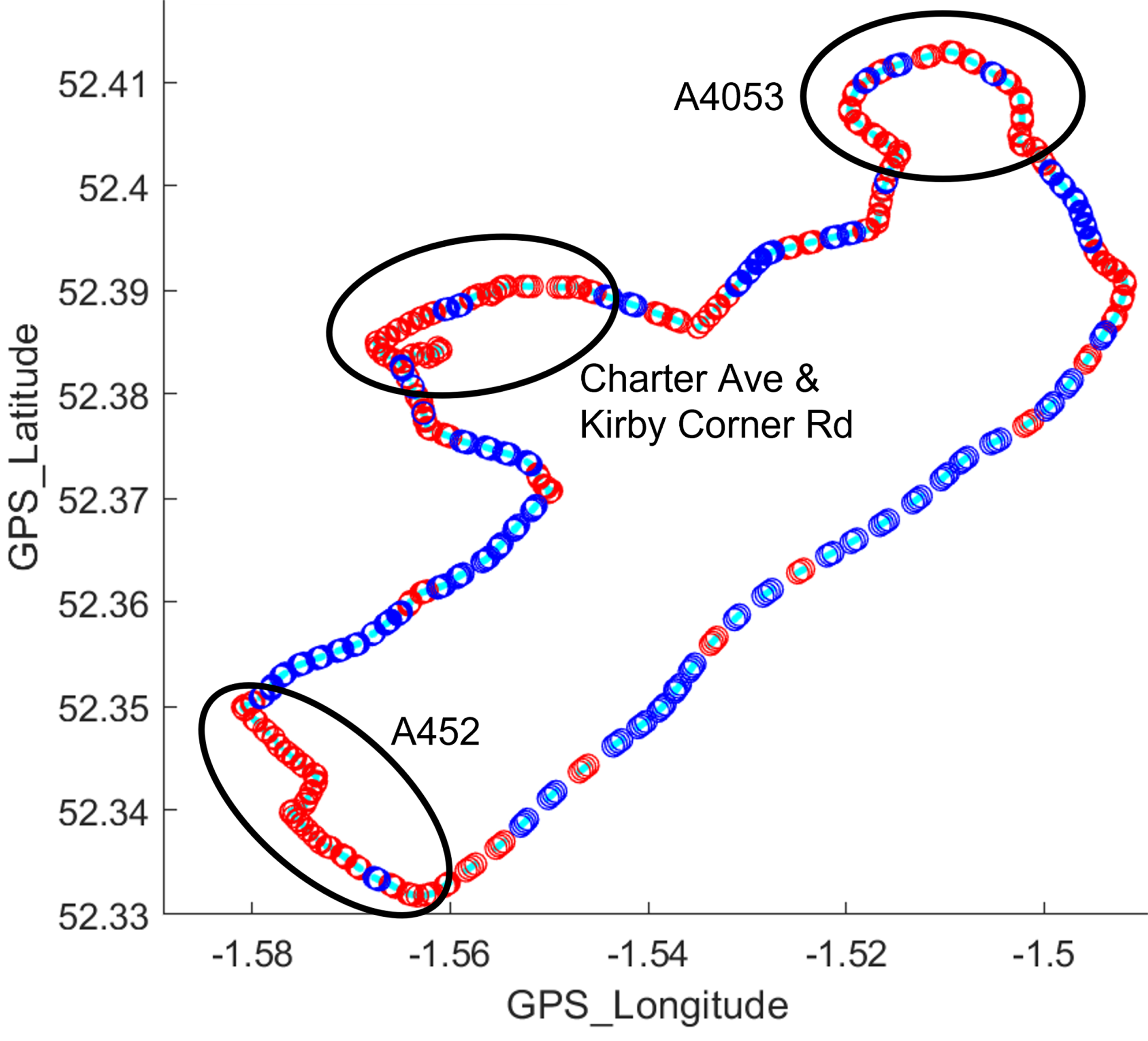}
\caption{P40.}\label{fig:WorkloadMap40}
\end{subfigure}
\begin{subfigure}[t]{0.492\linewidth}
\centering
\includegraphics[width=4.45cm,height=4.45cm]{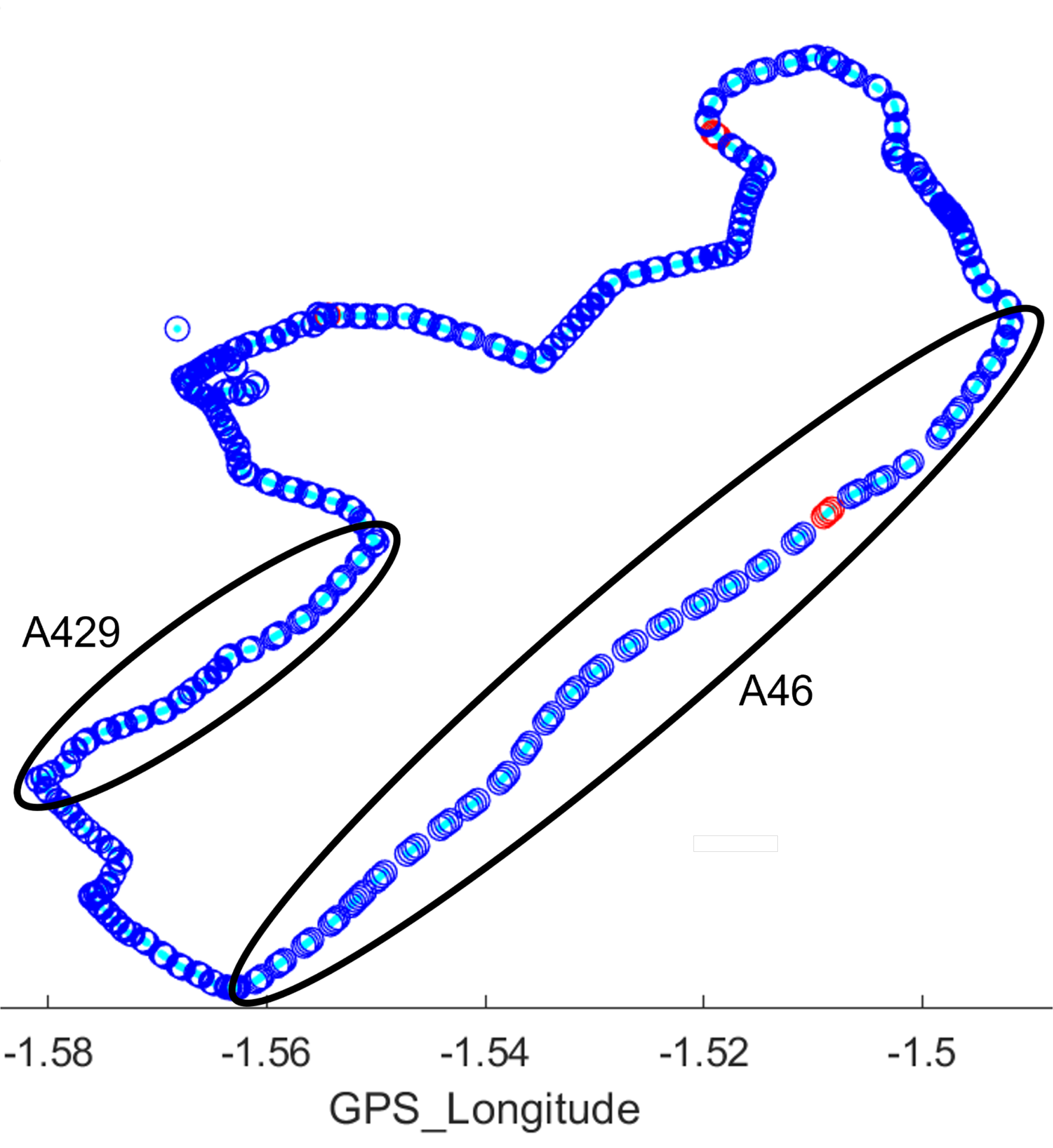}
\caption{P18.}\label{fig:WorkloadMap18}
\end{subfigure}                
\caption{Workload levels for two participants. Red, blue and cyan circles are high IWL, low IWL and request/prompt events, respectively.}
\label{fig:WorkloadMap}
\end{figure}

To demonstrate the differences between drivers belonging to distinct AWPs (e.g. low and high), the time/location of instantaneous workloads and issued requests (visual prompts) overlaid on the experimental route map for P40 with $\text{AWP} = H$ and P18 with $\text{AWP} = L$ are shown in Figure \ref{fig:WorkloadMap}. Here, P40 and P18 are ID codes assigned to the anonymous participants. In this case, the urban parts of the route caused high workload for P40 but not for P18, which emphasises the importance of AWP and the individuality of drivers. This highlights that these factors can be more influential than trivial environmental factors such as the road-type. Both participants indicated low workload for long stretches of the main road sections, i.e. A429 and A46, where the driving task is not demanding.

\vspace{-3mm}
\subsection{AWP and Driving Style (RQ2.1)}
To investigate potential links between the reported (subjective) driving style of each participant and the three recognised AWPs, we first cluster the DSQ data (see Section \ref{sec:collectedmeasures}) with hierarchical clustering. From the dendrogram, the resultant maximum number of clusters is four and 19 indices~\cite{charrad2014nbclust} propose three clusters as the best number. These three clusters are marked in Figure \ref{fig:WorkloadDSQ}, which shows that drivers of different AWP can belong to each of the three DSQ data clusters (i.e. driving styles). Hence, it can be deduced that there is no clear correlation between self-reported driving style from the DSQ data and AWPs. It is noted that Gaussian mixture model suggested 10 clusters and K-means produced an inconclusive number using the Bayesian information criterion and distortion\textunderscore fK measures, respectively \cite{mouselimis2022clusterr}. Such discrepancy between the outcomes of the three applied clustering techniques indicate that there is no clear separation between clusters of DSQ factors.

\begin{figure*}[htbp!]
	\centering    
	\includegraphics[width=2\columnwidth]{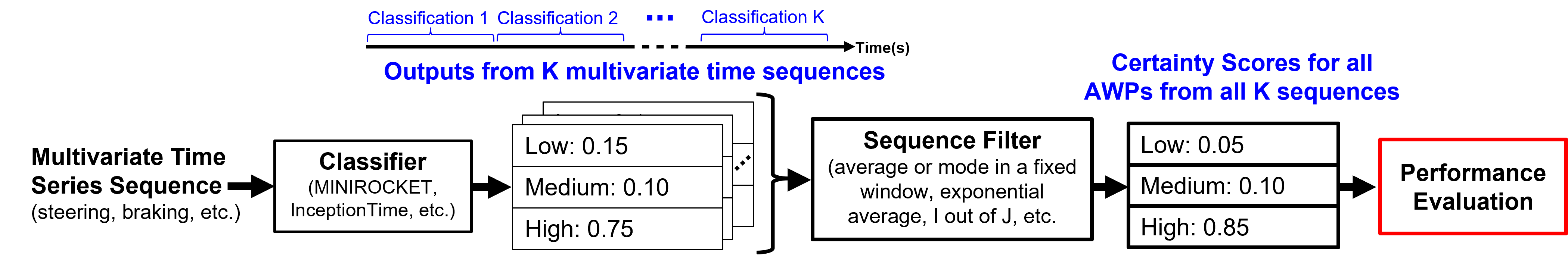}
	\caption[Framework for classifying drivers into average workload profiles using a sequence filter.]{Framework for classifying drivers into average workload profiles using a sequence filter.}
	\label{fig:OnRoad_EvalFrame}
\end{figure*}
\subsection{Discussion on Video Analysis and AWP}
Examining the reported data revealed that urban roads, often with high number of road users and diverse infrastructure, usually cause higher IWL compared with less demanding settings, such as main roads. This confirms findings in related studies on workload, e.g.~\cite{cnossen2000strategic,schneegass2013data}, and drivers' ability to safely perform secondary tasks on motorways~\cite{reimer2012field,mehler2012sensitivity}. The video analysis showed that vehicle control on junctions and monitoring the behaviour of vehicles in front are amongst the main contributors to driver's high instantaneous workload. Unlike prior work, this study connects automated (i.e. PDT-based) self-evaluated workload to environmental factors. It illustrates that a driver workload estimation solution for adaptive HMI can greatly benefit from access to measurements of contributing environmental factors such as road type (e.g. from landmark or road type identification system to recognise junctions or motorways, and mapping information) and other road users (e.g. from proximity sensors and pedestrian detection). Other unobserved factors (e.g. weather and badly maintained roads) can similarly induce frequent high IWL.


The frequency of reporting a high IWL drastically varied between participants, albeit all driving the same route. This is perhaps expected given drivers' individuality, e.g. considering fluctuations found in physiological workload measures~\cite{solovey2014classifying} and existing research into personalised workload prediction~\cite{yi2019personalized}. However, it is valuable to confirm here that such findings apply to drivers' subjective workload data. Three groups of drivers (i.e. AWPs) are identified, namely low, medium and high, emphasising the diverse capabilities among drivers (e.g. based on driving experience and familiarity with the route). It is imperative that knowledge of drivers' AWP can guide the IWL inference, e.g. it will bias the estimator towards low IWL values for drivers with $\text{AWP} = L$. It is noted that drivers' AWP can change over time (e.g. due to gained driving experience), hence regularly  predicting the drivers' AWP from accessible data (e.g. driving performance signals) can be desired, especially if results are to be employed by an adaptive HMI solution. This is an alternative to determining a driver's AWP from large amounts of historical data. 

The lack of correlation between objective driving style and AWP is in line with \cite{stahl2014correlations}; this however can still be due to the unsuitability of the standard DSQ \cite{french1993decision} for this purpose.

\section{Inference of Average Workload Profile}\label{sec:AWPestimation}
Here we address the problem of inferring the driver's AWP from driving performance data, i.e. RQ2.2 in Section \ref{sec:problemstatement}.  This is based on the premise that such data (e.g. when analysed over a finite time window) encapsulates the participant's driving experience and style depending on the perceived complexity of the driving task, which in turn directly influences or even dictates their AWP. In \cite{solovey2014classifying}, for instance, this point is highlighted by demonstrating that drivers react differently to the same external demand, indicating distinct workload management capabilities that can be captured in a short period of driving. The proposed approach is depicted in Figure \ref{fig:OnRoad_EvalFrame}.
\vspace{-3mm}
\subsection{Overall Classification Framework}
Matching a driver to one of the three identified profiles $\text{AWP}\in\{L,M,H\}$ is treated as a standard classification problem using supervised learning from the selected example CAN-bus signals in Table \ref{table:CAN_Bus_OnRoad} (excluding GPS) and their rate of change (except for the two steering wheel signals). They are all labelled with the corresponding AWP as per the participant's LWR in Figure \ref{fig:WorkloadDSQ}. Given the importance of harnessing temporal information/correlations and dependency between the various data streams to categorise a driver, the input to the recognition algorithm is a multivariate time series. Thus, this is a Time Series Classification (TSC) task, which is an alternative to manually engineering features from segments of data followed by a traditional classifier such as SVM and random forests.

Two possible TSC techniques, InceptionTime \cite{ismail2020inceptiontime} and MINIROCKET \cite{dempster2021minirocket}, are examined in this work. Each produces the confidence score in the driver's AWP for an input Multivariate Time-series Sequence (MTS). The choice is based on the classification algorithms' state-of-the-art accuracy, computational efficiency (e.g. a fraction of the computational cost of other deep learning techniques) and ability to extract non-linear discriminative features that are time-invariant; see \cite{ismail2020inceptiontime} and \cite{dempster2021minirocket} for an overview and benchmarking of various TSC approaches.  Whilst InceptionTime uses an ensemble of (five) deep convolutional neural network models (each with randomly initialised weights and created by cascading multiple Inception modules), MINIROCKET transforms the input time series with random convolutional kernels and applies a linear classifier to the transformed features. The same architecture and parameters as described in ~\cite{ismail2020inceptiontime,dempster2021minirocket} are used here.

A major challenge of effectively employing TSC in this work is the limited available data. For the $n^\text{th}$ participant, the data comprises $K_{n}$ time series sequences, each of length $L$: $\mathcal{D}^{(n)}= \left\{(X_1,Y_n),(X_2,Y_n),...(X_{K_{n}},Y_n)\right\}$ such that $X_i \in \mathbb{R}^{Q \times L}$ is the MTS, $Q=12$ is the number of considered signals (including their rates of change) and $Y_n \in \{L, M, H\}$ is the driver's AWP. The total dataset from the conducted experiment as in Section \ref{sec:experimentalstudy} is: $\mathcal{D} = \bigcup_{n=1}^{N}\mathcal{D}^{(n)}$ for all $N=24$ participants. Hence, the common TSC strategy of utilising longer MTSs (i.e. beyond a few minutes) is not viable here since it severely limits the training (and testing) data size, as experimental drive per participant is $\approx40$ minutes.

We therefore propose applying TSC multiple (successive) times, each on a short time series sequence, during a journey. A means, dubbed sequence filter, is then applied to combine the results from these multiple classifications. It can be a moving average, a majority voting scheme which requires a decision based on the result from each MTS
or even a more complex scheme such as a tracker of the classifier output. Here, we only investigate a Simple Moving Average (SMA); an exponential MA with a reasonable gain had a similar impact as the SMA and other techniques can be considered in future work. It is noted that not taking a decision on the inferred AWP prior to combining all the recognition results (e.g. as with SMA) is expected to better capture classification certainty over time. The category with the maximum certainty, i.e. Maximum a Posteriori (MAP) estimate, is admitted as the driver's AWP when assessing the classification performance.  

The performance evaluation block is added to Figure \ref{fig:OnRoad_EvalFrame} to highlight the need to interpret the TSC results in the context of the goal of this categorising task and how AWP information is used, i.e. facilitate personalised IWL estimation for adaptive HMI applications. For example, falsely classifying a driver with medium AWP as high AWP is more acceptable than classifying them as low AWP; this is discussed further below.

\begin{figure}[b]
	\centering    
	\includegraphics[width=1\columnwidth]{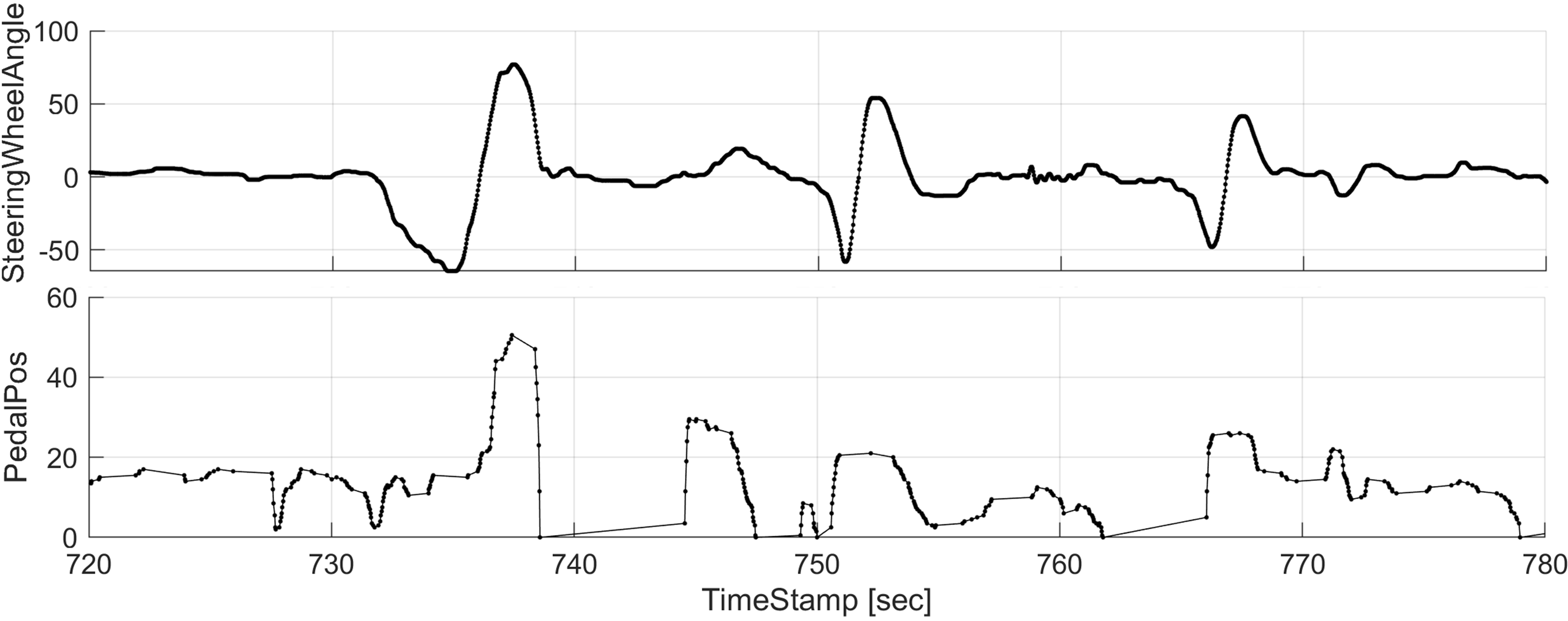}
	\caption{Two example raw CAN-bus signals for P40 within 1 minute time window; solid line is linear interpolation.}
	\label{fig:dataExample}
\end{figure}
\vspace{-4mm}
\subsection{Data Pre-Processing and Training Strategy}
Two example raw CAN-bus signals are displayed in Figure \ref{fig:dataExample} to show the asynchronous nature of the considered data streams, each with their own rate. Consequently, several steps were taken to prepare the data for TSC algorithms, since they expect the input (i.e. MTS) to be of fixed length and all its entries to be synchronised. Due to the nature of the AWP inference task (i.e. determining a driver's long-term workload capabilities), such  pre-processing (including buffering) of the received data and any resultant delays can be tolerated since no (near) real-time updates of the driver's AWP is required here contrary to the IWL estimation in Section \ref{sec:workloadestimation}. 
First, a linear interpolation was used to fill in for the missing values of the inherently asynchronous and irregularly spaced CAN-bus data streams. 
Second, the data set was downsampled by a factor of ten to have MTS of manageable size (raw signals are recorded at a rate of $\approx 200$ Hz). 
Third, various MTS periods between 5 and 90 seconds were defined for the time series sequences' length, i.e., $L \in \{100, 200, 400, 600, 1200, 1800\}$ corresponding to duration of $T_{\text{Seq}} =\{5,10,20,30,60,90\}$ seconds. Four, the data were scaled using the RobustScaler provided by the scikit-learn library. Finally, the data were split randomly into a training and test set with a ratio of $80\%$ to $20\%$, respectively.

\vspace{-3mm}
\subsection{Results and Discussion on AWP Inference}
Table \ref{table:OnRoad_Alg_Comp_Driver} depicts the classification accuracy and $F_{1}$ score without and with the sequence filter SMA. It clearly demonstrates the benefits of combining recognition results from multiple sequences. SMA consistently improves performance for both TSC models, e.g. by up to $\approx40\%$. It is noted that with the sequence filter we have only $N=24$ classifications in comparison to significantly more results (depending on $T_{\text{Seq}}$) for no SMA. This table also indicates that the quality of the classifications generally improves as the MTS duration increases; it then stagnates or declines after it exceeds 20 seconds for InceptionTime. This could be explained by the trade-off of increasing the sequence's length which provides more observable patterns at the expense of reducing the number of time series sequences available for training. Irrespective of the MTS length, MINIROCKET persistently delivers a superior TSC performance, 
which can be attributed to its ability to generalise better compared to InceptionTime. We stress that any other applicable sequence classifier can be assessed within the presented AWP inference framework.

\begin{table}
	\caption[$F_{1}$ score and accuracy for classification of drivers into Average workload profiles.]{$F_{1}$ score and accuracy of drivers classification into AWPs with and without a sequence filter for various durations $T_{\text{Seq}}$.}
	\centering
	\label{table:OnRoad_Alg_Comp_Driver}
	\begin{tabular}{l c c c  c c}
		\toprule
		\toprule

		Algorithm & $T_{\text{Seq}}$ & $F_{1}$score & Accuracy  & $F_{1}$score  & Accuracy  \\ 
		\midrule
		\multicolumn{2}{c}{}&\multicolumn{2}{c}{\textbf{w/o Sequence Filter}}& \multicolumn{2}{c}{\textbf{w/t Sequence Filter}} \\
		\midrule
		\multirow{6}{*}{\footnotesize{MINIROCKET}} & 5 s & 0.59 & 0.6 & 0.87 & 0.88 \\

		& 10 s & 0.59 & 0.61 & 0.83 & 0.83 \\
		
		& 20 s & 0.64 & 0.65 & 0.92 & 0.92 \\
		
		& 30 s & 0.61 & 0.63 & 0.77 & 0.79 \\
		
		& 60 s & 0.64 & 0.65 & 0.81 & 0.83 \\
		
		& 90 s & 0.63 & 0.65 & 0.92 & 0.92 \\
		
		\midrule
		
		\multirow{5}{*}{InceptionTime} & 5 s & 0.48 & 0.5 & 0.38 & 0.5 \\
		
		& 10 s & 0.51 & 0.53 & 0.67 & 0.71  \\
		
		& 20 s & 0.48 & 0.49 & 0.7 & 0.71 \\
		
		& 30 s & 0.5 & 0.56 & 0.52 & 0.58 \\
		
		& 60 s & 0.46 & 0.52 & 0.5 & 0.58 \\
		
		& 90 s & 0.46 & 0.5 & 0.45 & 0.5 \\
		\bottomrule
		\bottomrule
	\end{tabular}
\end{table}

\begin{figure}[b] 
	\centering    
	\includegraphics[width=8.5cm,height=6cm]{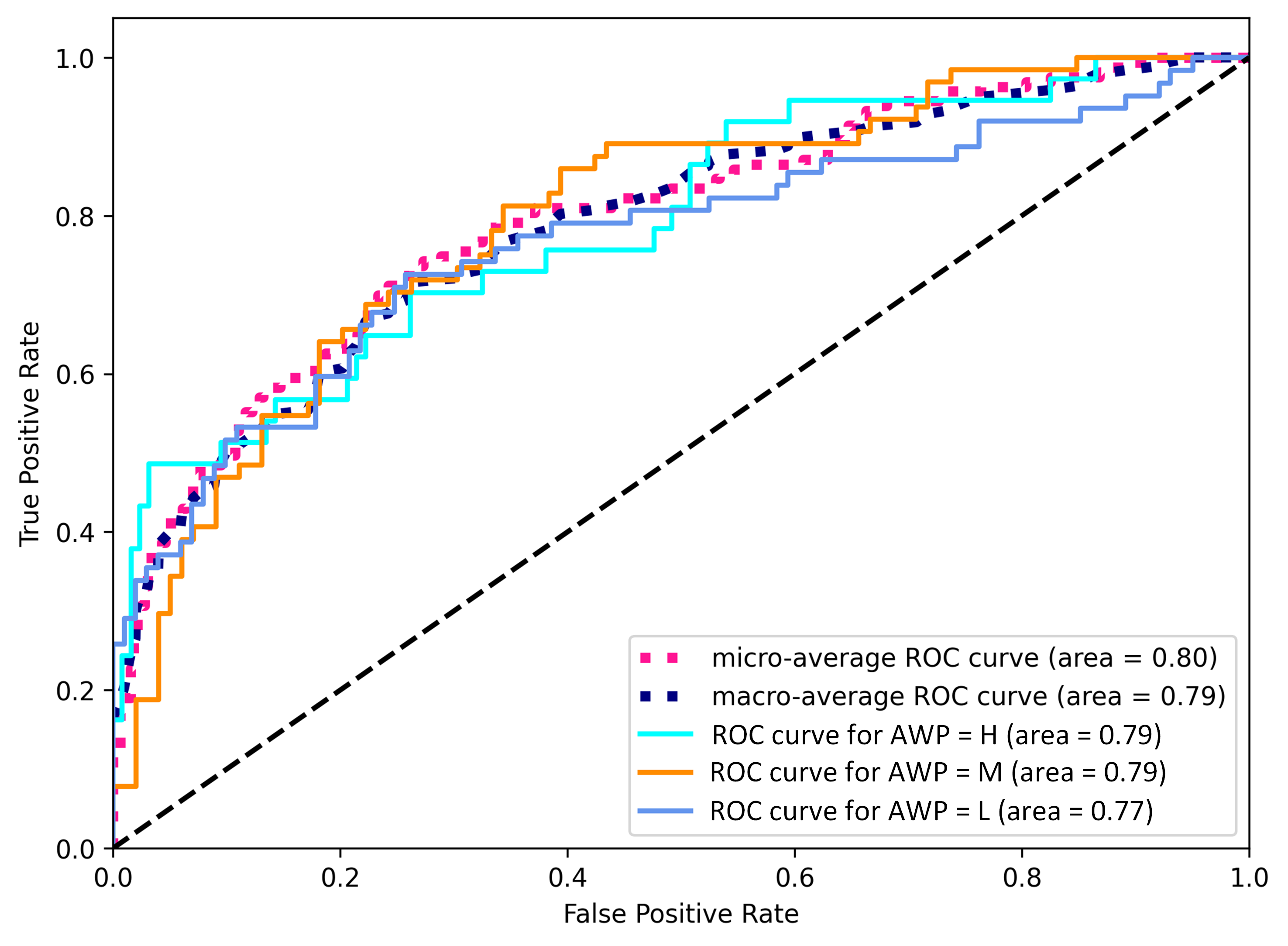}
	\caption[ROC curves for MINIROCKET with sequence length 20 s.]{ROC curves for MINIROCKET with sequence length 20 s. 
	}
	\label{fig:ROC_400}
\end{figure}

The MINIROCKET Receiver Operating Characteristic (ROC) and its area, i.e. Area Under the Curve (AUC), for the three AWP categories are shown in Figure \ref{fig:ROC_400} for $T_{\text{Seq}}=20$ seconds. It demonstrates the effectiveness of this TSC approach, where specific performance can be attained by the suitable choice of a certainty threshold value (instead of the MAP decision criterion). Interestingly, achieving the highest AUC for the high average workload profile is particularly desirable in the context of adaptive HMI since a classification of a driver within this class into a lower AWP (i.e. they are more available to interact with the HMI and/or receive information) can be considered more detrimental for driving safety and user satisfaction/acceptance than the contrary situation.  

The MINIROCKET confusion matrix for $T_{\text{Seq}}=\{20, 90\}$ seconds (identical for both) with the sequence filter is displayed in Table \ref{table:OnRoad_Driver_ConMat}. It can be seen that only two drivers (one with low and one with high AWP) are incorrectly categorised with a medium AWP, i.e. success rate $\approx92\%$. Additionally, it can be argued that such misclassifications can have a limited impact on personalisation for adaptive HMI (e.g. personalised IWL estimation) compared with placing a driver with a low AWP into the high category and vice versa.     

\begin{table}
	\caption[Confusion matrix showing the classification of drivers into average workload profiles using MINIROCKET with a sequence period of either 20 or 90 seconds.]{Confusion matrix showing the MINIROCKET performance  with a sequence period of either 20 or 90 seconds. 
	}
	\centering
	\label{table:OnRoad_Driver_ConMat}
	\begin{tabular}{l c c c c}
		\toprule
		 &  & \multicolumn{3}{c}{Predicted class} \\ 
		\cmidrule{3-5}
		 &  & Low  & Medium & High  \\ 
		\midrule
		\multirow{3}{*}{True class} & Low & 7 & 1 & 0  \\
		
		& Medium & 0 & 10 & 0  \\
		
		& High & 0 & 1 & 5 \\

		\bottomrule
	\end{tabular}
\end{table}



The explored AWP classification approach, comprising the MINIROCKET algorithm and a sequence filter, showed high performance, for instance 92\% accuracy in predicting the drivers' AWP. This is particularly notable since: 1) the input signals are only a few widely available driving performance data streams; 2) classifiers run on relatively short data segments; and 3) applied algorithms (namely MINIROCKET) have remarkably low computational resource requirements (i.e. are suitable for the automotive environment). Nonetheless, it is reasonable to argue that processing more vehicle signals (e.g. lane deviation and proximity to nearby vehicles) and even physiological measures (e.g. eye gaze and heart rate), and collecting larger data sets (e.g. permit using longer MTS) 
can further improve the classifier performance and generalisability, thereby facilitating a more reliable and personalised driver workload monitoring system. 

However, utilising standard classification performance metrics (e.g. accuracy and $F_{1}$ score) might not adequately capture the AWP classification efficacy without considering an adaptive HMI system's requirements. For instance, while inferring a lower than the true driver AWP category can have adverse effects on the user experience and safety, the opposite reduces an adaptive HMI's number of interactions. Thus, other performance metrics that weigh the outcomes for each class differently (e.g. based on the importance of its True Positives TPs and False Positives FPs) might be needed (e.g. skewed or weighted $F_{1}$ scores); formulating these appropriately should be investigated in future research. Next, we present an IWL estimator that can exploit AWP classifications results to deliver a personalised inference.


\section{Proposed Instantaneous Workload Estimator}  \label{sec:workloadestimation}
Here we describe and evaluate a simple Bayesian filtering approach to estimate driver's IWL
from asynchronous driving performance data streams, i.e. address RQ3 in Section \ref{sec:problemstatement}. Adapting the inference model is also discussed, e.g. incorporating AWP, road type information, and new data sources.
\vspace{-3mm}
\subsection{Bayesian Filtering Approach}
Let $y_k \in \mathbb{R}^{Q_n}$ be the observations (i.e. values of $Q_k$ CAN-bus signals) at time instant $t_k$; $y_{1:k}=\{y_{1},y_{2},...,y_{k}\}$ are the successive measurements collected at $\{t_{1},t_{2},...,t_{k}\}$ and are available at $t_k$. From Bayes' rule \cite{sarkka2013bayesian}, the posterior of the sought latent state $x_k$ (i.e. the driver's instantaneous workload) at $t_k$ based on $y_{1:k}$ can be expressed by: 
\begin{align} 
 &p(x_k|y_{1:k}) \propto p(y_k|x_k)p(x_k | y_{1:k-1})  \nonumber \\
 & ~= \textstyle p(y_k|x_k)\sum_{x_{k-1}\in\{L,H\}} p(x_k|x_{k-1})p(x_{k-1}|y_{1:k-1}),
 \label{eq:Bayes}
\end{align}
under the Markov assumption $p(x_k|x_{1:k-1}) = p(x_k|x_{k-1})$ and since $x_k\in\{L,H\}$  is a discrete variable, i.e. for low and high IWLs. It follows from the ``predict'' step with marginalisation: 
\begin{equation}
  p(x_k | y_{1:k-1})=\textstyle \int_{x_{k-1}} p(x_k|x_{k-1})p(x_{k-1}|y_{1:k-1})dx_{k-1}. \nonumber
\end{equation}

Since $p(x_{k-1}|y_{1:k-1})$ pertains to the previous time step $t_{k-1}$ and is available at $t_{k}$, computing $p(x_k|y_{1:k})$ in (\ref{eq:Bayes}) entails defining: i) dynamical model of the latent state evolution over time $x_{k}|x_{k-1}\sim p(x_k|x_{k-1})$ ; and ii) observation likelihood $y_{k}|x_{k}\sim p(y_k|x_k)$ which maps the measurements (e.g. CAN-bus data) to the state (i.e. IWL). Both are defined next.
\begin{figure}[!t]
	\centering    
	\includegraphics[width=0.75\columnwidth]{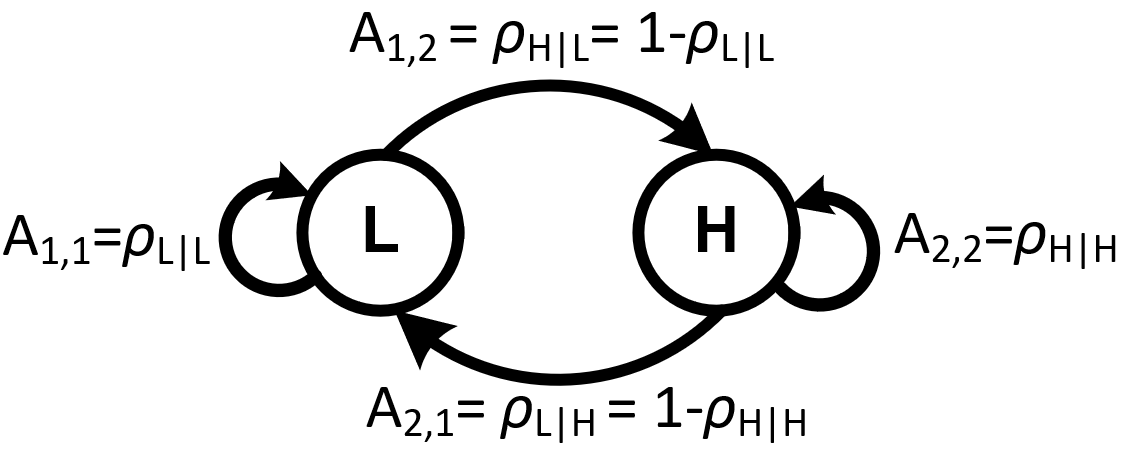}
	\caption{Markov chain model of IWL with transition matrix entries.}
	\label{fig:HMM}
\end{figure}
\begin{figure*}[htbp!]
	\centering    
	\includegraphics[width=2\columnwidth]{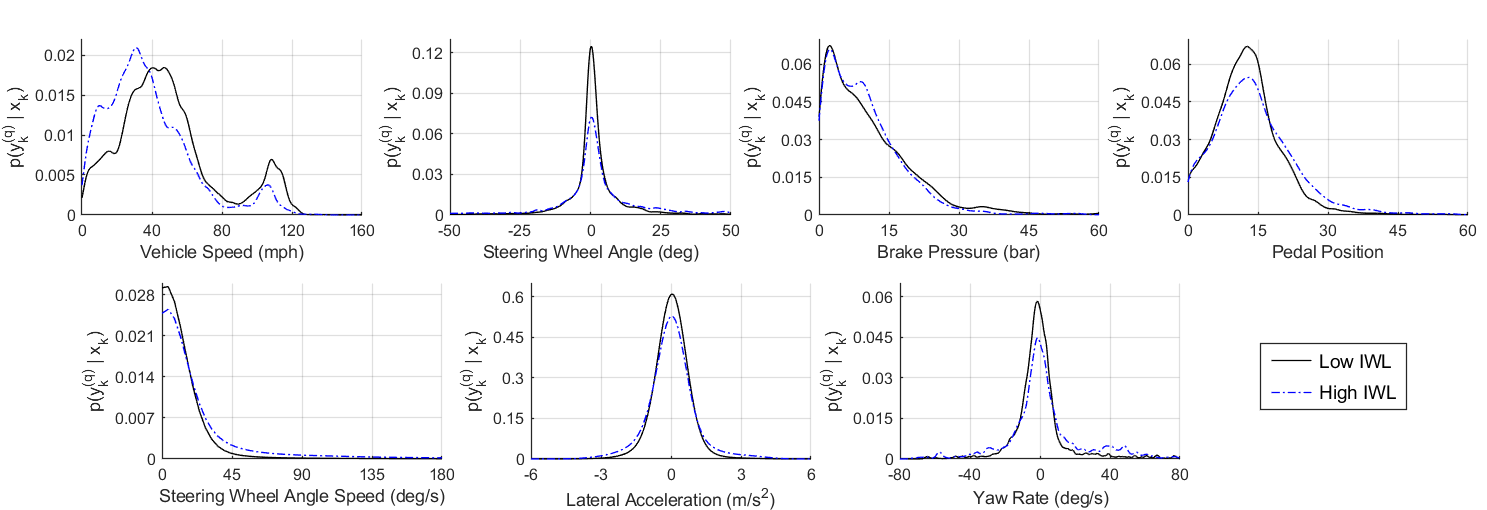}
	\caption{Univariate observation likelihoods $p(y^{(q)}_k|x_k)$ in (\ref{eq:likelihoodequation}) for $x_k\in\{L,H\}$ for the $Q=7$ used CAN-bus signals.}
	\label{fig:likelihoodExample}
\end{figure*}
\subsubsection{Dynamical Model}
We propose using the first-order Markov Chain (MC) in Figure \ref{fig:HMM} as a simple model for $p(x_k|x_{k-1})$. It is described by its transition matrix $A_k$, which can change over time, albeit not frequently (e.g. to adapt to a situation such as approaching a junction, and/or to personalise for a specific driver's AWP). Its entries\footnote{$A_{i,j}$ is the for the entry in the $i^{\text{th}}$  row and $j^{\text{th}}$ column of matrix $A$.} are:  
\begin{itemize}
    \item $A_{1,1,k}=\rho_{L|L,k}$: probability of low IWL remaining low
    \item $A_{1,2,k}=\rho_{H|L,k}$: probability of low IWL becoming high
    \item $A_{2,1,k}=\rho_{L|H,k}$: probability of high IWL becoming low
    \item $A_{2,2,k}=\rho_{H|H,k}$: probability of high IWL remaining high
\end{itemize}
such that $\rho_{L|L,k}+\rho_{H|L,k}=1$ and $\rho_{L|H,k}+\rho_{H|H,k}=1$ since $\forall i, \sum_{j} A_{i,j} =1$. These model parameters, which we only need to set two of them, are physically meaningful and represent prior knowledge on how we expect workload to change over time. For instance, while it is valid to assume that a low IWL can become high swiftly (e.g. due to environmental factors that induce a high IWL such as a front car suddenly braking), it is unrealistic for a high IWL to transition to low as rapidly. This implies that we should have $\rho_{H|L,k}>\rho_{L|H,k}$.
\subsubsection{Observation Likelihood} This likelihood is learnt for both possible scenarios, i.e. $p(y_k|x_k=H)$ and $p(y_k|x_k=L)$, from the available CAN-bus data of the on-road study, labelled by the corresponding reported IWL (i.e. around the time instants when requests to press the finger-worn button are issued and press events, if any). Each of the driving performance data streams are assumed to be independent conditioned on IWL, which is reasonable since different factors that can instigate a high IWL impact different signals. For example, a car braking at the front can affect pedal position but not necessarily lateral acceleration. We can then write:
\begin{equation}
    p(y_k|x_k) = \prod_{q=1}^{Q_k} p(y^{(q)}_k|x_k),
    \label{eq:likelihoodequation}
\end{equation}
where $y^{(q)}_k$ is for the $q^{th}$ data source (e.g. one of the CAN-bus signals in Table \ref{table:CAN_Bus_OnRoad}) and $Q_k$ is the total number of streams present  at $t_k$. This is owed to the asynchronous nature of the CAN-bus signals (e.g. see example signals in Figure \ref{fig:dataExample}). Here, the Kernel Density Estimate (KDE) \cite{bishopbook} is utilised to learn the univariate observation likelihood for each of the considered data streams from labelled data. This low-complexity non-parametric technique enables approximating Probability Density Functions (PDFs) of arbitrary shapes from data. Example observation likelihoods obtained from all $24$ participants are shown in Figure \ref{fig:likelihoodExample}, with noticeable  differences (albeit subtle at times and mostly distinguishable in the tail distributions for some signals) between the low and high workload scenarios for all considered $7$ data sources. For example, larger changes in the steering wheel angles are more likely to be observed in high IWL situations. PDFs from two or more signals, e.g. to capture possible correlations between them as with the bivariate likelihood of vehicle speed and braking in \cite{semmens2019now}, can be similarly obtained with multivariate KDE and employed within a suitably modified (\ref{eq:likelihoodequation}). We also recall that the seven CAN-bus signals in Figure \ref{fig:likelihoodExample} and Table \ref{table:CAN_Bus_OnRoad} were chosen to illustrate the IWL estimation capability and other streams (e.g. longitudinal acceleration, etc.) can be incorporated into \eqref{eq:likelihoodequation}.

\subsubsection{Bayesian Recursion}
With the above dynamical and observation models, the driver's IWL at $t_k$ can be \textit{sequentially} estimated from (\ref{eq:Bayes}) using the Bayesian filter (BF):  
\begin{align}\label{eq:Bayesiantrecursion}
\centering
& p(x_k=i|y_{1:k}) \approx \pi_{i,k}=\frac{\hat \pi_{i,k}}{\sum_{j} \hat \pi_{j,k}},~~~ i,j\in\{L,H\} \\
&\hat \pi_{L,k} = \ell_{L,k}\left[\rho_{L|L,k} \pi_{L,k-1} + (1-\rho_{H|H,k}) \pi_{H,k-1} \right],\nonumber \\
& \hat \pi_{H,k}= \ell_{H,k}\left[(1-\rho_{L|L,k}) \pi_{L,k-1}+\rho_{H|H,k} \pi_{H,k-1}\right], \nonumber
\end{align}
where $\ell_{L,k} = p(y_k|x_k = L)$ and $\ell_{H,k}= p(y_k|x_k = H)$ are the observation likelihoods.

\subsubsection{Computational and Implementation Complexities} 
A key advantage of the used Bayesian filter is its simplicity and low computational complexity; it is a closed-form analytical solution of the Bayesian recursion equation. In fact, implementing (\ref{eq:Bayesiantrecursion}) entails only a few add-multiply operations and the univariate likelihoods of (\ref{eq:likelihoodequation}) can be applied as lookup tables; thus the proposed inference is amenable to real-time implementation and supports producing estimates at relatively high rates (e.g. matching the incoming data frequencies), whilst demanding remarkably low computational resources.

\subsubsection{Adaptability of the IWL Estimator}
\begin{figure}[!t]
	\centering    
	\includegraphics[width=0.75\columnwidth]{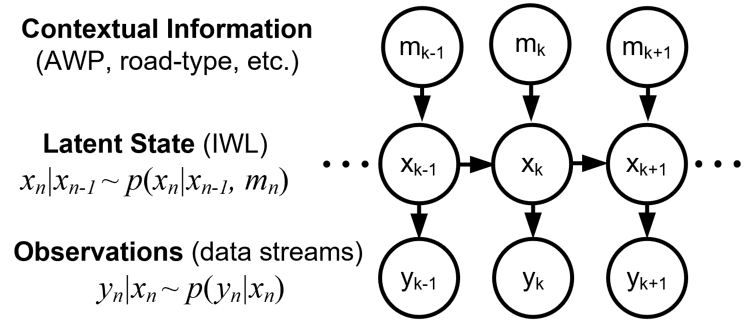}
	\caption{System graphical representation where contextual information $m_k$ (e.g. AWP or road-type) can drive the state transition.}
	\label{fig:systemgraph}
\end{figure}

Adapting or personalising the estimator can be achieved by adjusting: a) either the dynamical model state transition matrix $A$ (e.g. to facilitate a faster transition to and retaining a high  IWL via larger $\rho_{H|L}$ and $\rho_{H|H}$ values for drivers with high AWP); and/or b) the observation likelihood function. The former approach can be interpreted as conditioning the state transition on external information $m_k$ at time $t_k$ (e.g. AWP and road-type) such that $x_k|x_{k-1}\sim p(x_k|x_{k-1},m_k)$. 
 This revised system model is shown in Figure \ref{fig:systemgraph}. 
Most importantly, this does not involve retraining the inference model (albeit the need to choose an appropriate $A_k$ based on $m_k$). Furthermore, incorporating data from new sources (e.g. other raw or processed CAN-bus signals, sensory measurements from other vehicle sensors such as proximity to nearby vehicles, and even biometrics data) can be handled within the adopted Bayesian formulation by including its univariate likelihood in (\ref{eq:likelihoodequation}). 

Therefore, the proposed estimator of the driver instantaneous workload level not only has low computational and implementational complexities, but also it can be easily personalised. Examples of such adaptations are presented below.

\vspace{-5mm}
\subsection{IWL Estimation Results}
The CAN-bus signals in Table \ref{table:CAN_Bus_OnRoad} (except GPS) are used here by the proposed IWL estimator, which is applied directly on the asynchronous data streams without any pre-processing steps (e.g. interpolation) and the typical associated latencies. At each time instant, the Bayesian filter updates the workload posterior distribution in (\ref{eq:Bayesiantrecursion}) from the available data sources at $t_k$. Thus, it can  produce estimates at the same rate as the input signals. In the performance evaluation below, which is based on over 339,000 high IWL and 953,000 low IWL events (i.e. around and between the requests to press the button), the BF observation likelihoods are learnt with KDE from data of all participants except the driver being tested. It is noted that these PDFs do not noticeably differ from the ones in Figure \ref{fig:likelihoodExample}. Whereas, the model transition matrix $A_k$ entries in Table \ref{table:transitionMatrix} are chosen intuitively and from (minimal) manual tuning.
\begin{table}
   \caption{Various transition matrices with their diagonal entries.}
	\centering
	\label{table:transitionMatrix}
	\begin{tabular}{l c c c c c}
		\toprule
	     Probability & $A_{k}^{\text{Standard}}$&$A_{k}^{(\text{H})}$ & $A_{k}^{(\text{Ha})}$ & $A_{k}^{(\text{La})}$ & $A_{k}^{(\text{L})}$
	     \\ 
	    \midrule
	    $\rho_{L|L}$ & 0.8 & 0.4 & 0.7 & 0.75 & 0.9 
	    \\
		$\rho_{H|H}$ & 0.92 & 0.98 & 0.92 & 0.8 & 0.8\\
		\bottomrule
	\end{tabular}
\end{table}

First, we benchmark the introduced Bayesian filtering technique with $A_{k}^{(\text{Standard})}$ 
against the following two methods:
\begin{itemize}
\item
Rule-based technique in \cite{wright2017intelligent}: utilises prescribed membership functions and entails defining features such as cruising state derived from the vehicle speed.
\item
SVM classifier: uses standard SVM as in \cite{semmens2019now,yi2019new} with a Gaussian radial basis function and treats determining IWL as a binary classification task. 
\item 
LSTM: adopts an off-the-shelf long short-term sequence classifier as in \cite{Goodfellow-et-al-2016}. A sequence length $L=10$ is employed for this task, based on experimenting with various $L$ values and increasing the length (e.g. to 20) did not result in an improved performance.
\end{itemize}
Since these three techniques expect the input vectors to be of the same length, CAN-bus signals had to be interpolated to cope with missing data. For both SVM and LSTM, a stratified sampling strategy is applied in order to account for the skewed class proportions when training from all available data, except that pertaining to the tested participant (i.e. as with attaining the BP observation likelihoods). Hence, the same evaluation process is performed for each driver and the average performance is computed. The results are displayed in Table \ref{table:IWLBenchmarking} such that the true positives are for high IWL and true negatives are for low IWL. Threshold that leads to the highest $F_{1}$ score is used for deciding IWL from the probabilities/scores of the BF and rule-based methods.

Table \ref{table:IWLBenchmarking} shows that BF delivers the best performance and the SVM-based approach exhibits the poorest overall results. This can be attributed to the probabilistic and sequential nature (i.e. intrinsically propagates and combines estimation certainty over time) of the presented Bayesian filtering approach to IWL estimation. It outperforms the LSTM-based sequence classifier, which is a more suitable standard classifier for use with the processed sequence data compared to more conventional techniques such as SVM. 
Nonetheless, more sophisticated machine learning techniques for time series modelling such as transformers \cite{tf2017nips} and more elaborate Recurrent Neural Networks (RNNs) are not explored further here since these methods cannot be readily employed for multiple highly asynchronous data streams with output expected in (near) real-time unlike with AWP in Section \ref{sec:AWPestimation}. Adapting them to accommodate missing data is still an open research question and is out of the scope of this paper.

\begin{table}
	\caption{Average accuracy, recall, precision and $F_{1}$ score for the rule-based, SVM, LSTM and proposed Bayesian filtering techniques.}
	\centering
	\label{table:IWLBenchmarking}
	\begin{tabular}{l c c c c}
		\toprule
	     Estimator & Accuracy &Recall & Precision &  $F_{1}$ score 
	     \\ 
	    \midrule
	    SVM & 0.59 & 0.16 & 0.26 & 0.12 
	    \\
LSTM & 0.66 & 0.30 & 0.41 & 0.30 
	    \\
		Rule-based & 0.43 &0.72& 0.37& 0.35
		\\
        Proposed BF & 0.72 &0.85 & 0.62 & 0.67 
        \\
		\bottomrule
	\end{tabular}
\end{table}

Next, we illustrate how the introduced Bayesian filtering can be adapted by adjusting its transition matrix $A_{k}$ as per:
\begin{enumerate}
    \item \textbf{Road-type} (e.g. from navigation service, road signs identification, etc.): four such types are assumed based on their potential to induce a high IWL, see heatmap in Figure \ref{fig:HeatMap}, namely junctions with $A_{k}^{(\text{H})}$, urban road with $A_{k}^{(\text{Ha})}$, country road with $A_{k}^{(\text{La})}$  and motorway with $A_{k}^{(\text{L})}$. 
    \item \textbf{Average Workload Profile} (i.e. from the inference framework in Section \ref{sec:AWPestimation}): $A_{k}^{(\text{L})}$, $A_{k}^{(\text{Standard})}$  and $A_{k}^{(\text{H})}$ for drivers with low, medium and high AWP, respectively. In this case, the IWL estimates, in near real-time, incorporate the AWP inference results. 
\end{enumerate}

\begin{figure}[t] 
	\centering    
	\includegraphics[width=\columnwidth]{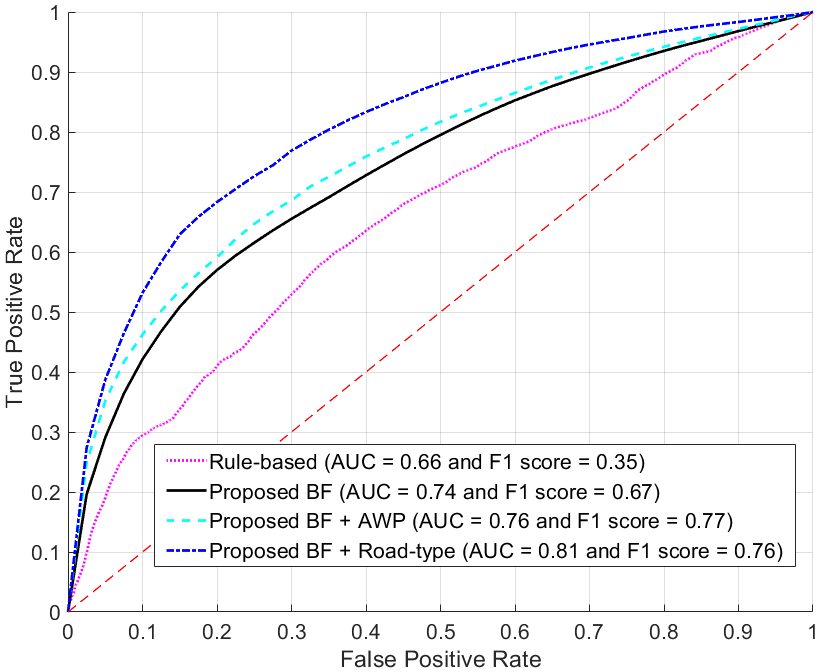}
	\caption{ROC curves, AUCs and $F_{1}$ scores of the rule-based method in \cite{wright2017intelligent} and the various considered versions of the introduced BF.}
	\label{fig:IWLAdapted}
\end{figure}

The ROC, area under its curve (AUC), and $F_{1}$ scores of these two adaptations, BF using a fixed $A_{k}^{\text{(Standard)}}$ and rule-based method in \cite{wright2017intelligent} are depicted in Figure \ref{fig:IWLAdapted} for all $24$ participants. The figure illustrates that significant improvements in the workload estimation can be achieved by suitably personalising the inference model and confirms the importance of incorporating contextual information (e.g. AWP and road-type) as noted in \cite{semmens2019now}. Whilst using road-type led to the biggest enhancement in AUC and the ROC curve, utilising AWP led to a modest increase of $\approx15\%$ in the $F_{1}$ score. This highlights that certain environmental factors, such as road-type, can have more pronounced effects on drivers in terms of inducing high IWL (i.e. true positives) as per the video analysis in Table \ref{table:Video_Workload}. Conversely, participants with a low AWP and substantially high low workload ratio in (\ref{eq:LWR}), i.e. tend to often report a low workload level  as seen with P18 in Figure \ref{fig:WorkloadMap18}, are less exposed to such factors. For instance, $F_{1}$ score for driver P18 remarkably increased from $0.23$ to $0.95$ when including AWP information. It is noted that in Figure \ref{fig:IWLAdapted} the results of the 10 drivers with medium workload profiles did not change since $A_{k}^{\text{Standard}}$ was adopted for them. Adapting the IWL inference model based on several simultaneous contextual information (e.g. AWP, road-type and others) would require devising rules to reflect their collective impact on the driver's workload, which is not treated here. 
\vspace{-3mm}
\subsection{Discussion on IWL Estimation}
The introduced simple and computationally efficient Bayesian filtering approach delivers a reasonable, overall IWL estimation performance, particularly when considering that only seven CAN-bus signals were used. Utilising other data streams, including other raw or processed CAN-bus signals and even biometric data if available, is expected to enhance the inference accuracy. This can be readily accomplished with BF via updating the observation likelihood in (\ref{eq:likelihoodequation}). 

On the other hand, advanced machine learning algorithms, such as time series classifiers (e.g. transformers and carefully designed RNN networks), can be applied to determine the driver's IWL and have the potential to achieve superior results compared to BF. This is provided that they can be used with multivariate asynchronous time series signals and that sufficiently large training datasets are available. The LSTM performance in Table \ref{table:IWLBenchmarking} might have been restricted by the size of the available training dataset, which is limited compared to other application scenarios with speech and text data. The adopted Bayesian filtering however has several clear advantages compared with such classifiers, namely it: i) naturally handles asynchronous data arriving at different rates (i.e. no pre-processing steps are required) and is amenable to real-time implementations; ii) has low training data requirements with reduced risks of over-fitting, iii) is fully explainable and causes of failures can be quickly identified; iv) can be easily adapted or personalised based on additional information (e.g. driver profile) without requiring an extensive training dataset specific to the new scenario (e.g. a profile); and v) is  inherently sequential and combines/integrates results across time (i.e. no need for a sequence filter).  We emphasise that the choice of the model parameters here, i.e. matrix $A_k$, did not involve extensive fine tuning, which indicates that BF is a robust means for instantaneous workload estimation. Nevertheless, an invalid choice of priors in $A_k$ and/or configuring the observation likelihoods based on unrepresentative data can severely degrade the BF performance. Learning $A_k$ from the data for a chosen observation likelihood function $p(y_k|x_k)$, e.g. by maximising the resultant likelihood $p(y_{1:k}|x_{1:k})$ for entire journeys, in fact led to a reduced average inference accuracy here; this parameterisation strategy might be better suited for learning and applying a specific $A_k$ for each driver.

Finally, more complex transition models can be tackled by the closed-form BF from equation (\ref{eq:Bayes}), for instance for IWL with more than two possible values based on a new labelling scheme of the subjective workload that takes into account the duration between a visual request/prompt and the user pressing the button. Other continuous state models for $x_k$ can also be explored within the introduced Bayesian tracking formulation, e.g. the IWL is a continuous random variable that can take any value within a defined range.


\section{Conclusion} \label{sec:conclusions}
This is the first study to collect drivers' self-evaluated workload through a modified PDT in naturalistic driving scenarios and identify external factors that induce high workload levels.  
While junctions and monitoring other road users (e.g. cars in front) are associated with high workload levels, driving on a motorway often led to low workload. Data analysis additionally revealed three average (long-term) workload profiles, i.e. low, medium and high, amongst the drivers.
This highlights drivers' individuality and the driving environment's influence on workload assessment, reiterating the necessity for personalised workload estimation. Subsequently, time series classification within a supervised learning framework was first shown to be capable of accurately matching a driver to one of the aforementioned workload profiles from the driving performance data. A simple and flexible Bayesian filtering approach was then introduced for estimating the driver's instantaneous workload from a number of selected asynchronous CAN-bus signals. Its ability to effectively leverage contextual information (such as road type and driver profile) to achieve enhanced estimation performance was demonstrated, i.e. personalised inference. Whilst only seven example driving performance data streams were considered here, more sources can be used by the proposed instantaneous workload estimator. 

This study serves as an impetus to motivate further research into robust workload estimation for adaptive HMI that can fuse asynchronous as well as heterogeneous data streams (if/when available) and leverage contextual information for personalised workload inference. Bayesian filtering methods is shown here to be a good candidate for the IWL estimation task and more extensive benchmarking against other existing as well as continuously emerging machine learning algorithms (e.g. classifiers) from larger on-road datasets should be considered in future work.  


\appendices
\section*{Acknowledgment}
The authors would like to thank Jaguar Land Rover for
funding this work under the CAPE agreement.
\ifCLASSOPTIONcaptionsoff
  \newpage
\fi
\bibliographystyle{IEEEtran}
\bibliography{IEEEfull,references}








\end{document}